\documentclass[]{pasj02} 
\usepackage[switch,mathlines]{lineno} % add line number to manuscript
\usepackage{natbib}
\jyear{2026}
\Received{2026/04/01}%{yyyy/mm/dd}
\Accepted{2026/06/17}%{yyyy/mm/dd}
\Published{}%{yyyy/mm/dd}
\usepackage{amsmath}
\usepackage{url}
\usepackage{pdflscape}
\usepackage{longtable}
\usepackage{arydshln}
\defcitealias{Shoshi_2025_oph}{I}

\begin{document}
\title{ALMA 2D super-resolution imaging survey of Ophiuchus Class~I/flat spectrum/II disks. II. Statistical analysis of stellar and disk properties}
\author{
Ayumu \textsc{Shoshi},
\altaffilmark{1,2}
\altemailmark\orcid{0000-0001-6580-6038} 
\email{shoshi.ayumu.660@s.kyushu-u.ac.jp} 
Masayuki \textsc{Yamaguchi},
\altaffilmark{3,4}
\altemailmark\orcid{0000-0002-8185-9882} 
%\email{myamaguchi@asiaa.sinica.edu.tw} 
Takayuki \textsc{Muto}
\altaffilmark{5}
\altemailmark%\orcid{}
%\email{muto@cc.kogakuin.ac.jp}
Naomi \textsc{Hirano},
\altaffilmark{2}
\altemailmark\orcid{0000-0001-9304-7884}
%\email{hirano@asiaa.sinica.edu.tw}
Ryohei \textsc{Kawabe},
\altaffilmark{4, 6}
\altemailmark\orcid{0000-0002-8049-7525}
%\email{ryo.kawabe32@gmail.com}
Takashi \textsc{Tsukagoshi},
\altaffilmark{7}
\altemailmark\orcid{0000-0002-6034-2892}
%\email{takashi.tsukagoshi.astro@gmail.com}
Shu \textsc{Ishibashi},
\altaffilmark{8}
\altemailmark\orcid{0009-0001-1889-9216}
%\email{k8174036@kadai.jp}
and
Masahiro \textsc{N. Machida}
\altaffilmark{3}
\altemailmark\orcid{0000-0002-0963-0872}
%\email{machida.masahiro.018@m.kyushu-u.ac.jp}
}
\altaffiltext{1}{Department of Earth and Planetary Sciences, Graduate School of Science, Kyushu University, 744 Motooka, Nishi-ku, Fukuoka 819-0395, Japan}
\altaffiltext{2}{Academia Sinica Institute of Astronomy and Astrophysics, 11F of ASMA Building, No.1, Sec. 4, Roosevelt Rd, Taipei 106216, Taiwan}
\altaffiltext{3}{Department of Earth and Planetary Sciences, Faculty of Science, Kyushu University, 744 Motooka, Nishi-ku, Fukuoka 819-0395, Japan}
\altaffiltext{4}{National Astronomical Observatory of Japan, 2-21-1 Osawa, Mitaka, Tokyo 181-8588, Japan}
\altaffiltext{5}{Division of Liberal Arts, Kogakuin University, 1-24-2 Nishi-Shinjuku, Shinjuku-ku, Tokyo 163-8677, Japan}
\altaffiltext{6}{Department of Astronomy, School of Science, The Graduate University for Advanced Studies (SOKENDAI), Osawa, Mitaka, Tokyo 181-8588, Japan}
\altaffiltext{7}{Faculty of Engineering, Ashikaga University, Ohmae-cho 268-1, Ashikaga, Tochigi 326-8558, Japan}
\altaffiltext{8}{Department of Physics and Astronomy, Graduate School of Science and Engineering, Kagoshima University, 1-21-35 Korimoto, Kagoshima, Kagoshima 890-0065, Japan}

\maketitle

%%% Abstract %%%
\begin{abstract}
We present a statistical study of stellar and dust disk properties for young stellar objects in the Ophiuchus star-forming region.
Building on our previous paper \citep[][]{Shoshi_2025_oph}, which applied two-dimensional super-resolution imaging with \texttt{PRIISM} to ALMA archival Band~6 continuum data and spatially resolved 78 disks, we analyze a sample of 67 systems with robust dust-radius measurements.
We combine stellar parameters from the literature, including bolometric temperature $T_{\rm bol}$, stellar mass $M_\ast$, and mass accretion rate $\dot{M}_{\rm acc}$, with disk parameters derived from the super-resolution images, including inclination $i_{\rm disk}$, millimeter luminosity $L_{\rm mm}$, and dust radius $R_{95\%}$.
We quantify pairwise correlations and compare their behavior across evolutionary stages (Class~I/FS and Class~II) and between disks with and without detectable substructures.
We identify substructure dependencies in $L_{\rm mm}$ and $R_{95\%}$, indicating that substructures tend to be found preferentially in relatively massive and extended disks.
Moreover, we find a tight size--luminosity relation between $R_{95\%}$ and $L_{\rm mm}$.
In particular, only Class~II disks with substructures exhibit a steeper scaling, $R_{95\%}\propto L_{\rm mm}^{0.8}$, while the other subsamples are broadly consistent with $R_{95\%}\propto L_{\rm mm}^{0.4\text{--}0.5}$.
This behavior is qualitatively consistent with disk evolution models in which disks with planet-induced pressure bumps follow a steeper size-luminosity relation than smooth disks.
Overall, our results suggest that disk substructures play an important role in shaping the evolution of dust and global disk properties, while providing empirical constraints on accretion, dust trapping, and possible gravitational instability in young disks.
\end{abstract}
%\pagewiselinenumbers 

%%% Introduction %%%
\section{Introduction}\label{sec:intro}
Protoplanetary disks around young stellar objects (YSOs) are the sites of planet formation, and observations with the Atacama Large Millimeter/submillimeter Array (ALMA) have revealed their detailed structures with unprecedented clarity over the past decade.
In particular, the discovery of multiple rings and gaps in the HL~Tau disk provided a striking early demonstration of disk substructures \citep[e.g.,][]{ALMA_2015,Yen_2016}. 
This result spurred the idea that planet-disk interactions could generate such structures \citep[e.g.,][]{Dong_2015,Kanagawa_2016,Zhang_2018}.
While substructures may also arise from other mechanisms, local enhancements in gas pressure associated with rings or spirals can act as dust traps, efficiently concentrating large grains and thereby promoting the formation of planetesimals and, ultimately, planets \citep[e.g.,][]{Youdin_2005,Johansen_2007,Johansen_2009}.
Disk substructures are therefore regarded as both potential signposts of young planets and key environments that facilitate planet formation.

Motivated by their importance, several ALMA surveys in nearby star-forming regions have been conducted to detect and characterize disk substructures systematically \citep[e.g.,][]{Ansdell_2016,Barenfeld_2017,Cieza_2019,Long_2019,Cazzoletti_2019,Villenave_2021}.
The Cycle~4 large project, the Disk Substructures at High Angular Resolution Project \citep[DSHARP;][]{Andrews_2018_dsharp}, observed nearby Class~II disks at a resolution of $\sim$5\,au and demonstrated that rings, gaps, and spirals are ubiquitous in these systems.
Likewise, the Ophiuchus Disk Survey Employing ALMA project \citep[ODISEA;][]{Cieza_2021} observed Class~II disks in the Ophiuchus molecular cloud at resolutions of 3-5\,au and found ring-gap structures in essentially all targeted systems, indicating that prominent substructures are already well developed by the Class~II stage.

Prior to the Class~II stage, ALMA observations have also reported disk substructures in a handful of Class~0/I systems \citep[e.g.,][]{Sheehan_2017,Sheehan_2018,Sheehan_2020,Sai_2020,Maureira_2024,Shoshi_2026}.
In contrast, the Cycle~7 Large Program Early Planet Formation in Embedded Disks \citep[eDisk;][]{Ohashi_2023} performed a systematic survey of Class~0/I disks at $\sim$7\,au resolution and found that prominent substructures are comparatively rare at these early stages relative to Class~II disks.
This apparent contrast motivates an investigation into the evolutionary stage at which substructures become common and how their emergence depends on stellar and disk properties.

To make progress on this question, in \citet{Shoshi_2025_oph} (hereafter Paper~\citetalias{Shoshi_2025_oph}), we applied two-dimensional super-resolution imaging based on sparse modeling with \texttt{PRIISM}\footnote{\texttt{PRIISM} (Python Module for Radio Interferometry Imaging with Sparse Modeling) is a public software package for imaging ALMA observations based on the sparse modeling technique, and it is available at $\langle$\url{https://github.com/tnakazato/priism}$\rangle$.} \citep{Nakazato_2020,Nakazato_2020b} to archival ALMA Band 6 continuum observations of Class~I, flat-spectrum (FS), and Class~II disks in the Ophiuchus molecular cloud \citep[at a distance of $\sim$140\,pc;][]{Ortiz_2018,Gaia_2023}.
This imaging approach has been increasingly used in recent ALMA disk studies \citep[e.g.,][]{Yamaguchi_2020,Yamaguchi_2021,Yamaguchi_2025,Shoshi_2025_iras,Chou_2025,Shoshi_2026}, and its ability to deliver high-fidelity images has been demonstrated by \citet{Yamaguchi_2024} and Paper~\citetalias{Shoshi_2025_oph}.
As a result, we achieved spatial resolutions of a few au and resolved all targets, enabling a characterization of disk morphologies across the sample.
We identified previously unreported substructures in 15 disks and, by combining our sample with the eDisk survey, demonstrated for the first time that prominent substructures are restricted to systems with bolometric temperatures exceeding 200-300\,K and dust disk radii larger than $\sim$30\,au.
These bolometric temperatures correspond to ages of $\sim$0.2-0.4\,Myr after the onset of star formation \citep{Evans_2009}, indicating that prominent substructures are already in place during the main mass accretion phase.
A similar trend was also subsequently found independently by \citet{Hsieh_2025}.

In Paper~\citetalias{Shoshi_2025_oph}, our analysis was primarily focused on disk substructures, and the statistical characterization was limited to the distributions of disk inclination and dust radius. 
In this second paper of the series, we extend that work by carrying out a more detailed statistical study of both stellar and spatially resolved dust disk properties for the same Ophiuchus sample. 
We compile stellar quantities from the literature and disk quantities derived from the super-resolution images. 
Using this catalog, we investigate how distributions and correlations of these stellar and disk properties depend not only on the evolutionary stage (Class~I/FS versus Class~II), but also on the presence or absence of disk substructures.
This paper is organized as follows. 
In section~\ref{sec:statistics}, we describe the sample selection and the compilation of stellar and disk properties. 
Section~\ref{sec:results} presents an overview of the statistical results for the stellar and dust disk parameters. 
In section~\ref{sec:discussion}, we discuss the correlations involving mass accretion rate and dust mass and compare them between smooth and substructured disks as well as between different evolutionary stages. 
Finally, section~\ref{sec:summary} summarizes our main conclusions.

%%% Sample Selection %%%
\section{Sample and Analysis Methods}\label{sec:statistics}
\subsection{Sample selection}\label{subsec:selection}
In Paper~\citetalias{Shoshi_2025_oph}, we used the same ALMA archival data (project 2016.1.00545.S; PI: Lucas A. Cieza) as the ODISEA survey \citep[][]{Cieza_2019}.
The observed targets were 147 young stellar objects in the Ophiuchus molecular cloud, selected from the ``Cores to Disks'' (c2d) Spitzer Legacy Program sample.
This sample includes Class~I and flat-spectrum (FS) sources with $[K]-[24] > 6.75$\,mag and Class~II sources brighter than 10\,mag in the $K$ band.
After applying self-calibration and super-resolution imaging with \texttt{PRIISM}, the final sample used in Paper~\citetalias{Shoshi_2025_oph} consisted of 78 disks, associated with 70 single stars and, for 8 multiple systems, the brightest 1.3\,mm continuum component in each system.
The remaining 69 sources in the original ALMA dataset were excluded from the analysis because either their signal-to-noise ratios (S/Ns) were too low to reveal any significant structure in the CLEAN images, or the cross-validation used to optimize the regularization hyperparameters in the PRIISM imaging did not converge to a stable solution (for details, see section~3 of Paper~\citetalias{Shoshi_2025_oph}).

For the statistical analysis presented in this study, we constructed a sample by carefully reselecting from the previous 78 disks, as described below.
We first exclude ISO-Oph~200, ISO-Oph~137, BBRCG~58, ISO-Oph~171, 2MASS~J16314457-2402129, and ISO-Oph~106 because the maximum baseline lengths were insufficient to constrain the radii of the six disks (see appendix~2.2 of Paper~\citetalias{Shoshi_2025_oph}).
Furthermore, to ensure that our measurement values are robust, we require that the diameter of the spatially resolved dust disks, defined as $2R_{95\%}$, where $R_{95\%}$ is the dust radius enclosing 95\% of the 1.3\,mm total flux density, be resolved by at least three times the major-axis full width half-maximum (FWHM) of the effective spatial resolution in the PRIISM images (see tables~2 and table~3 of Paper~\citetalias{Shoshi_2025_oph}).
We finally employ 67 systems listed in table~\ref{table:properties} as the targets for this statistical work.
Specifically, the two Class~I (ISO-Oph~37 and ISO-Oph~200), six Class~FS (ISO-Oph~46, ISO-Oph~107, ISO-Oph~132, BBRCG~58, ISO-Oph~171, and ISO-Oph~212), and three Class~II (ISO-Oph~39, ISO-Oph~72, and ISO-Oph~106) sources were excluded based on the resolution requirements mentioned above.
Note that most of the removed systems have total 1.3\,mm flux densities of $<20$\,mJy and dust disk radii of $<$20\,au, indicating that our sample for the statistical analysis preferentially includes brighter and larger disks and is therefore incomplete for faint and compact disks.

\subsection{Stellar and disk properties}\label{subsec:properties}
Table~\ref{table:properties} summarizes six stellar and dust disk properties for the analysis targets.
Stellar quantities such as the bolometric temperature $T_{\rm bol}$, stellar mass $M_\ast$, and mass accretion rate $\dot{M}_{\rm acc}$ are compiled from the literature (see the footnotes of table~\ref{table:properties}).
Dust disk quantities are derived from the PRIISM images of Paper~\citetalias{Shoshi_2025_oph}.
The inclination angles $i_{\rm disk}$ were measured from two-dimensional Gaussian fitting with the Markov chain Monte Carlo (MCMC) method for all disks, and, for 13 disks (WL~17, ISO-Oph~51, Elias~24, WSB~82, ISO-Oph~2, ISO-Oph~196, DoAr~44, ISO-Oph~17, SR~24S, RXJ1633.9-2442, IRAS~16201-2410, SR~13, and SR~4) with visually distinguishable ring structures, from an ellipse fit to the rings (see section~4.1 of Paper~\citetalias{Shoshi_2025_oph}).
We characterized the dust disk size using $R_{95\%}$, defined as the radius enclosing 95\% of the total 1.3\,mm flux density $F_\nu$. 
We further quantified the millimeter luminosity $L_{\rm mm}$ by $F_\nu$ scaled to a distance of 140\,pc.
The uncertainties in $L_{\rm mm}$ and $R_{95\%}$ include a 10\% absolute flux calibration uncertainty for ALMA observations and the standard deviation of the geometric mean of the effective spatial resolution in the PRIISM image, respectively (see Paper~\citetalias{Shoshi_2025_oph}).

For each pair among the six quantities ($T_{\rm bol}$, $M_\ast$, $\dot{M}_{\rm acc}$, $i_{\rm disk}$, $L_{\rm mm}$, and $R_{95\%}$), we estimate the Pearson correlation coefficient $\rho$ and the corresponding $p$-value for the full samples (all disks) and two subsamples (Class~I/FS disks and Class~II disks).
Note that we conservatively exclude ISO-Oph~36 and ISO-Oph~167 from statistical analyses involving $M_\ast$ because their stellar masses are high relative to their reported spectral types.
In estimating these coefficients, we use logarithmic values for all quantities except the inclination.
Throughout this work, we regard a correlation as statistically robust and significant when the $p$-value is less than 0.003, and as tentative when $0.003\leq p<0.05$. 
We also describe their correlation strength using $\lvert\rho\rvert$ as a guide, referring to $\lvert\rho\rvert<0.3$ as weak, $0.3\leq\lvert\rho\rvert<0.6$ as moderate, and $\lvert\rho\rvert\geq0.6$ as strong.
Using these significance and strength criteria, we identify correlations among these quantities and examine how their distributions depend on evolutionary stage and the presence or absence of dust disk substructures.

{\onecolumn
\begin{landscape}
\begin{longtable}[t]{lccccccccccl}
\caption{Properties of host stars and dust disks}
\label{table:properties}\\
\hline
Name & Class~& $d$ & SpT & $T_{\rm bol}$ & $M_\ast$ & $\log\,\dot{M}_{\rm acc}$ & $i_{\rm disk}$ & $L_{\rm mm}$ & $R_{95\%}$ & Category & References \\
 &  & pc &  & K & $M_\odot$ & $M_\odot\,{\rm yr}^{-1}$ & $^\circ$ & mJy & au &  &  \\
(1) & (2) & (3) & (4) & (5) & (6) & (7) & (8) & (9) & (10) & (11) & (12) \\
\hline
\endfirsthead
\multicolumn{12}{l}{(Continued)}\\
\hline
Name & Class~& $d$ & SpT & $T_{\rm bol}$ & $M_\ast$ & $\log\,\dot{M}_{\rm acc}$ & $i_{\rm disk}$ & $L_{\rm mm}$ & $R_{95\%}$ & Category & References \\
 &  & pc &  & K & $M_\odot$ & $M_\odot\,{\rm yr}^{-1}$ & $^\circ$ & mJy & au &  &  \\
(1) & (2) & (3) & (4) & (5) & (6) & (7) & (8) & (9) & (10) & (11) & (12) \\
\hline
\endhead
ISO-Oph~54 & I & 140.0 & M4 & 380 & 0.58 & --- & 31.6$^{+0.24}_{-0.21}$ & 95.8$\pm$9.6 & 123.7$\pm$12.5 & Single/Inflection & a, c, e, i, x, z \\
2MASS~J16214513-2342316 & I & 140.0 & --- & 240 & --- & --- & 80.2$^{+0.02}_{-0.02}$ & 38.5$\pm$3.9 & 85.9$\pm$3.8 & Single/Ring & a, i, x \\
WLY~2-63 & I & 140.0 & M6 & 270 & 0.65 & -7.30 & 46.9$^{+0.03}_{-0.03}$ & 353.5$\pm$35.4 & 79.0$\pm$5.8 & Single/Inflection & a, c, e, i, n, x \\
ISO-Oph~127 & I & 140.0 & --- & 630 & 0.48 & --- & 77.5$^{+0.02}_{-0.02}$ & 27.2$\pm$2.7 & 60.0$\pm$4.4 & Single/Ring & a, h, i, r, x \\
ISO-Oph~99 & I & 140.0 & --- & 97 & 0.23 & -6.12 & 75.8$^{+0.02}_{-0.02}$ & 52.5$\pm$5.3 & 51.1$\pm$4.4 & Single/Ring & a, i, l, s, x \\
ISO-Oph~165 & I & 140.0 & M2.5 & 320 & 0.3 & -8.90 & 73.6$^{+0.03}_{-0.03}$ & 37.4$\pm$3.7 & 43.3$\pm$3.7 & Single/Ring & a, d, i, r, x  \\
ISO-Oph~21 & I & 140.0 & M2 & 490 & --- & --- & 49.0$^{+0.04}_{-0.04}$ & 73.3$\pm$7.3 & 40.0$\pm$5.8 & Single/Smooth & a, i, t, x \\
2MASS~J16313679-2404200 & I & 140.0 & --- & 74 & --- & --- & 78.5$^{+0.05}_{-0.05}$ & 13.2$\pm$1.3 & 30.6$\pm$2.4 & Single/Smooth & a, i, x \\
2MASS~J16262548-2423015 & I & 140.0 & --- & 140 & --- & --- & 67.5$^{+0.02}_{-0.02}$ & 46.2$\pm$4.6 & 29.1$\pm$3.7 & Single/Smooth & a, i, x \\
ISO-Oph~170 & I & 140.0 & --- & 360 & --- & -11.07 & 71.4$^{+0.06}_{-0.05}$ & 12.6$\pm$1.3 & 26.7$\pm$3.6 & Single/Ring or Binary/Smooth? & a, h, i, o, x \\
2MASS~J16271643-2431145 & I & 140.0 & --- & 620 & --- & --- & 55.7$^{+0.13}_{-0.12}$ & 18.6$\pm$1.9 & 25.5$\pm$6.2 & Single/Smooth & a, i, x \\
2MASS~J16230544-2302566 & I & 140.0 & M3 & 790 & 0.40 & --- & 69.7$^{+0.39}_{-0.39}$ & 3.8$\pm$0.4 & 24.1$\pm$2.1 & Single/Smooth & a, e, i, r, x \\
WL~17 & I & 140.0 & M3 & 330 & 1.45 & -6.44 & 34.8$^{+0.80}_{-0.80}$ & 52.7$\pm$5.3 & 22.4$\pm$2.9 & Single/Ring & a, b, h, i, m, x \\
2MASS~J16313124-2426281 & F & 147.0 & K4 & 750 & 1.20 & --- & 86.0$^{+0.01}_{-0.01}$ & 51.2$\pm$5.1 & 144.5$\pm$4.6 & Single/Ring & a, i, j, x \\
2MASS~J16254662-2423361 & F & 140.0 & --- & 690 & --- & --- & 85.1$^{+0.02}_{-0.02}$ & 21.0$\pm$2.1 & 126.1$\pm$3.0 & Single/Ring & a, i, x \\
ISO-Oph~37 & F & 140.0 & K7 & 710 & 0.61 & -8.56 & 71.1$^{+0.01}_{-0.01}$ & 133.0$\pm$13.3 & 96.4$\pm$5.7 & Single/Ring & a, d, i, r, x  \\
ISO-Oph~94 & F & 140.0 & M1 & 770 & 0.36 & -9.67 & 79.2$^{+0.02}_{-0.02}$ & 34.6$\pm$3.5 & 69.1$\pm$4.6 & Single/Ring & a, e, h, i, r, x \\
2MASS~J16395292-2419314 & F & 140.0 & --- & 980 & --- & --- & 85.4$^{+0.04}_{-0.04}$ & 5.3$\pm$0.5 & 54.1$\pm$1.4 & Single/Ring & a, i, x \\
ISO-Oph~70 & F & 140.0 & M0 & 440 & --- & --- & 70.0$^{+0.03}_{-0.04}$ & 49.2$\pm$4.9 & 51.1$\pm$4.1 & Multiple/Ring & a, e, h, i, x \\
ISO-Oph~112 & F & 140.0 & M4 & 610 & 1.20 & -7.74 & 69.0$^{+0.03}_{-0.03}$ & 42.1$\pm$4.2 & 49.1$\pm$5.1 & Single/Inflection & a, d, e, h, i, x \\
ISO-Oph~93 & F & 140.0 & M3 & 380 & 0.31 & -8.49 & 71.9$^{+0.04}_{-0.04}$ & 19.6$\pm$2.0 & 41.8$\pm$4.4 & Single/Ring & a, e, h, i, x  \\
ISO-Oph~51 & F & 136.6 & M0 & 680 & 0.59 & -9.50 & 24.5$^{+0.70}_{-0.70}$ & 10.6$\pm$1.1 & 29.2$\pm$3.7 & Single/Ring & a, d, g, i, x \\
ISO-Oph~26 & F & 140.0 & M6 & 710 & 0.14 & -8.58 & 50.4$^{+0.06}_{-0.06}$ & 29.2$\pm$2.9 & 27.3$\pm$4.8 & Single/Smooth & a, d, h, i, x \\
ISO-Oph~167 & F & 140.0 & K6 & 570 & 3.80$^{\ddagger}$ & -6.43 & 48.0$^{+0.03}_{-0.03}$ & 74.5$\pm$7.5 & 24.2$\pm$4.4 & Multiple/Smooth & a, e, h, i, x \\
ISO-Oph~52 & F & 140.0 & M0 & --- & 0.17 & -8.42 & 65.7$^{+0.10}_{-0.10}$ & 8.6$\pm$0.9 & 23.7$\pm$3.7 & Single/Smooth & a, e, h, x \\
ISO-Oph~75 & F & 140.0 & M5 & 680 & 0.13 & -7.93 & 47.7$^{+0.25}_{-0.22}$ & 8.2$\pm$0.8 & 18.2$\pm$3.6 & Single/Smooth & a, e, h, i, x \\
ISO-Oph~129 & F & 140.0 & K7 & 660 & 0.61 & --- & 57.7$^{+0.03}_{-0.03}$ & 25.6$\pm$2.6 & 18.2$\pm$3.0 & Single/Smooth & a, e, h, i, r, x \\
ISO-Oph~95 & F & 140.0 & M4 & 640 & 0.95 & -9.24 & 38.0$^{+0.34}_{-0.39}$ & 14.5$\pm$1.4 & 16.4$\pm$4.1 & Single/Smooth & a, e, h, i, x \\
ISO-Oph~204 & F & 153.0 & M3 & 730 & --- & --- & 29.5$^{+0.07}_{-0.07}$ & 53.7$\pm$5.4 & 15.3$\pm$3.0 & Multiple/Smooth & a, g, i, u, x \\
ISO-Oph~59 & F & 140.0 & --- & 790 & 0.39 & -8.75 & 45.9$^{+0.10}_{-0.10}$ & 9.2$\pm$0.9 & 12.7$\pm$2.6 & Single/Smooth & a, h, i, x  \\
ISO-Oph~147 & F & 140.0 & K8 & 500 & 0.60 & -7.90 & 40.2$^{+0.18}_{-0.16}$ & 6.2$\pm$0.6 & 6.4$\pm$1.5 & Multiple/Smooth & a, d, i, x \\
Elias~27 & II & 110.1 & K8 & 820 & 0.63 & -7.14 & 56.2$^{+0.03}_{-0.03}$ & 161.5$\pm$16.2 & 178.9$\pm$7.4 & Single/Spiral & a, d, g, i, x \\
DoAr~25 & II & 138.2 & K6 & 1500  & 0.80 & -9.09 & 65.5$^{+0.03}_{-0.03}$ & 222.4$\pm$22.2 & 152.5$\pm$9.9 & Single/Inflection & a, d, g, i, x \\
Elias~24 & II & 139.3 & K5.5 & 980 & 1.10 & -6.34 & 22.7$^{+0.80}_{-0.80}$ & 337.3$\pm$33.7 & 132.6$\pm$7.3 & Single/Ring & a, d, g, i, x \\
WSB~82 & II & 145.8 & K4 & 1100  & 1.26 & --- & 50.9$^{+0.30}_{-0.30}$ & 130.5$\pm$13.1 & 118.1$\pm$10.9 & Single/Ring & a, g, i, k, x \\
ISO-Oph~2 & II & 134.3 & M0 & 1100  & 0.50 & -8.55 & 36.5$^{+1.00}_{-1.00}$ & 63.4$\pm$6.3 & 80.6$\pm$10.0 & Multiple/Ring & a, d, e, g, h, i, x \\
ISO-Oph~196 & II & 135.0 & M4.5  & 1200  & 0.22 & -7.95 & 36.1$^{+1.30}_{-1.30}$ & 80.2$\pm$8.0 & 70.0$\pm$3.0 & Single/Ring & a, d, g, i, x \\
DoAr~44 & II & 146.3 & K3 & 1200  & 1.40 & -8.20 & 23.3$^{+0.70}_{-0.70}$ & 81.2$\pm$8.1 & 67.9$\pm$6.5 & Single/Ring & a, b, e, i, x \\
ISO-Oph~17 & II & 140.0 & K8 & 290 & 0.69 & -7.63 & 40.5$^{+0.50}_{-0.50}$ & 167.6$\pm$16.8 & 63.7$\pm$6.0 & Single/Ring & a, d, i, x \\
RXJ1633.9-2442 & II & 143.8 & K7 & 1500  & 0.80 & -9.90 & 47.3$^{+0.30}_{-0.30}$ & 76.6$\pm$7.7 & 55.0$\pm$6.2 & Single/Ring & a, e, g, i, p, x \\
SR~24S$^\dagger$ & II & 100.4 & K2 & 840 & 1.50 & -7.47 & 47.8$^{+0.05}_{-0.05}$ & 93.3$\pm$9.3 & 53.6$\pm$4.7 & Multiple/Ring & a, c, f, g, i, x, y \\
Elias~20 & II & 137.5 & M0 & 990 & 0.88 & -6.68 & 51.5$^{+0.04}_{-0.04}$ & 89.2$\pm$8.9 & 51.8$\pm$3.9 & Single/Inflection & a, d, g, i, x \\
SR~20W & II & 146.8 & K5 & 1200  & 0.97 & --- & 70.5$^{+0.05}_{-0.05}$ & 25.4$\pm$2.5 & 45.4$\pm$5.4 & Single/Ring & a, d, g, i, x \\
IRAS16201-2410 & II & 156.6 & M0 & 1200  & 1.12 & --- & 49.9$^{+0.80}_{-0.80}$ & 51.4$\pm$5.1 & 41.8$\pm$4.3 & Single/Ring & a, e, g, k, i, x \\
SR~13 & II & 115.5 & M2 & 1300  & 0.37 & -8.70 & 31.0$^{+1.30}_{-1.30}$ & 40.2$\pm$4.0 & 32.6$\pm$3.5 & Multiple/Ring & a, d, g, h, i, x \\
SR~4 & II & 134.8 & K6 & 1100  & 0.80 & -7.03 & 23.9$^{+0.11}_{-0.13}$ & 60.2$\pm$6.0 & 31.5$\pm$5.3 & Single/Ring & a, d, g, i, x \\
DoAr~43 & II & 135.9 & K2 & 990 & 1.61 & --- & 74.7$^{+0.04}_{-0.04}$ & 13.7$\pm$1.4 & 30.9$\pm$2.2 & Multiple/Ring & a, e, g, i, v, x \\
ISO-Oph~105 & II & 134.6 & K7 & 1000  & 0.67 & -8.14 & 60.3$^{+0.06}_{-0.06}$ & 36.0$\pm$3.6 & 29.9$\pm$5.3 & Single/Smooth & a, d, g, i, x \\
WSB~52 & II & 135.3 & M0 & 1000  & 0.50 & -7.84 & 52.1$^{+0.04}_{-0.04}$ & 61.5$\pm$6.2 & 29.5$\pm$5.8 & Single/Smooth & a, d, g, i, x \\
DoAr~33 & II & 141.6 & K5 & 1600  & 0.98 & -9.60 & 43.4$^{+0.11}_{-0.11}$ & 33.3$\pm$3.3 & 27.1$\pm$5.0 & Single/Smooth & a, d, g, i, q, x \\
WSB~63 & II & 136.5 & M1.5 & 1400  & 0.39 & -9.30 & 73.4$^{+0.07}_{-0.06}$ & 11.6$\pm$1.2 & 26.6$\pm$3.1 & Single/Ring or Binary/Smooth? & a, d, g, i, q, x \\
ISO-Oph~117 & II & 140.7 & M3 & 830 & 0.29 & -8.23 & 37.4$^{+0.18}_{-0.19}$ & 21.2$\pm$2.1 & 24.3$\pm$4.6 & Single/Smooth & a, d, g, i, x  \\
WSB~19 & II & 142.0 & M3 & 1200  & 0.29 & --- & 41.5$^{+0.33}_{-0.34}$ & 11.1$\pm$1.1 & 24.0$\pm$5.5 & Multiple/Smooth & a, d, i, x \\
WSB~12 & II & 136.9 & K5 & 1400  & 1.30 & -8.00 & 54.9$^{+0.10}_{-0.09}$ & 27.4$\pm$2.7 & 21.9$\pm$4.8 & Multiple/Smooth & a, e, g, f, i, p, x \\
ISO-Oph~83 & II & 136.5 & K7 & 1300  & 0.72 & -8.13 & 49.1$^{+0.10}_{-0.10}$ & 14.3$\pm$1.4 & 21.3$\pm$3.5 & Single/Smooth & a, d, g, h, i, x \\
WSB~14 & II & 137.5 & --- & 1500  & --- & --- & 36.5$^{+0.18}_{-0.15}$ & 13.7$\pm$1.4 & 17.9$\pm$4.1 & Single/Smooth & a, g, i, x \\
ISO-Oph~163 & II & 139.5 & K5.5 & 1100  & 0.98 & -8.24 & 24.1$^{+0.14}_{-0.14}$ & 36.1$\pm$3.6 & 17.8$\pm$3.4 & Single/Smooth & a, d, g, i, x \\
WSB~67 & II & 141.1 & M3 & 1300  & 0.50 & --- & 43.0$^{+0.28}_{-0.28}$ & 6.7$\pm$0.7 & 17.6$\pm$3.2 & Single/Ring or Binary/Smooth? & a, f, g, i, x \\
SR~22 & II & 131.4 & M4 & 1400  & 0.22 & --- & 69.0$^{+0.05}_{-0.05}$ & 11.7$\pm$1.2 & 16.7$\pm$2.5 & Single/Smooth & a, d, g, i, x \\
DoAr~32 & II & 141.1 & K5 & 1400  & 0.98 & -9.70 & 33.8$^{+2.02}_{-1.85}$ & 3.4$\pm$0.3 & 13.6$\pm$3.0 & Single/Smooth & a, d, g, i, q, x \\
ISO-Oph~155 & II & 136.6 & K5.5 & 1200  & 1.14 & -7.26 & 35.5$^{+0.06}_{-0.06}$ & 25.1$\pm$2.5 & 12.7$\pm$3.2 & Single/Smooth & a, d, g, i, x \\
ISO-Oph~128 & II & 140.0 & M1.5  & 930 & 0.47 & -8.51 & 60.1$^{+0.30}_{-0.29}$ & 3.8$\pm$0.4 & 12.6$\pm$2.4 & Single/Smooth & a, d, g, i, x \\
ISO-Oph~62 & II & 136.7 & K7 & 1300  & 0.87 & -7.88 & 59.1$^{+0.03}_{-0.04}$ & 12.6$\pm$1.3 & 10.7$\pm$1.0 & Multiple/Smooth & a, d, i, x \\
ISO-Oph~36 & II & 139.2 & K0 & 1100  & 2.60$^{\ddagger}$ & -8.01 & 59.8$^{+0.03}_{-0.03}$ & 18.1$\pm$1.8 & 9.9$\pm$1.6 & Multiple/Smooth & a, d, g, i, x \\
ISO-Oph~20 & II & 135.3 & K6 & 1600  & 0.83 & -8.59 & 44.1$^{+0.13}_{-0.13}$ & 6.8$\pm$0.7 & 9.1$\pm$1.9 & Single/Smooth & a, d, g, i, x \\
ISO-Oph~116 & II & 137.2 & M0 & 1300  & 0.53 & -8.68 & 45.5$^{+0.18}_{-0.14}$ & 5.6$\pm$0.6 & 8.9$\pm$1.9 & Single/Smooth & a, d, g, i, x \\
2MASS~J16314457-2402129 & II & 132.0 & --- & 970 & --- & --- & 25.4$^{+0.67}_{-1.41}$ & 7.6$\pm$0.8 & 8.4$\pm$2.1 & Single/Smooth & a, g, i, x \\
\hline
\multicolumn{12}{p{\linewidth}}{\footnotesize
{\bf Column description:} 
(1) Source name.
(2) Classification for the host star mainly referred to \citet{Cieza_2019} and \citet{Williams_2019}.
(3) Distance of the host star adopted mainly from Gaia DR3 and Gaia DR2 parallax \citep{Gaia_2018,Gaia_2023}.
Most Class~I and FS sources, embedded in optically thick envelopes, lack individual distance estimates, so we adopt 140.0\,pc, the mean distance of Gaia DR2/DR3-detected systems in our sample.
(4) Spectral type of the host star.
(5) Bolometric temperature of the host star.
(6) Stellar mass.
(7) Mass accretion rate to the host star.
(8) Inclination angle of the dust disk at 1.3\,mm.
(9) Millimeter luminosity at the total flux at 1.3\,mm scaled at the distance of 140\,pc. 
An uncertainty includes a 10\% absolute calibration error for ALMA observations.
(10) Disk radius containing 95\% of the flux density, measured from the 1.3\,mm dust continuum emission via the curve-growth method.
The uncertainty is approximately equal to the square root of the effective spatial resolution in the PRIISM image.
(11) Disk categorization determined based on the intensity profile (see \S4.2 in Paper~\citetalias{Shoshi_2025_oph}).
A $\dagger$ symbol denotes that we updated the distance based on the Gaia DR3 parallax and revised the values of disk properties ($L_{\rm mm}$ and $R_{95\%}$) associated with the distance.
A $\ddagger$ symbol denotes that the stellar mass $M_\ast$ is high compared with the reported spectral type and is conservatively excluded from statistical analyses involving $M_\ast$.
}\\
\multicolumn{12}{p{\linewidth}}{\footnotesize
{\bf References.} 
a: \citet{Williams_2019},
b: \citet{Cieza_2019},
c: \citet{Cieza_2021},
d: \citet{Testi_2022},
e: \citet{vanderMarel_2021},
f: \citet{vanderMarel_2016},
g: \citet{Gaia_2023},
h: \citet{Manara_2015},
i: \citet{Dunham_2015},
j: \citet{Villenave_2022},
k: \citet{Michel_2021},
l: \citet{Nakamura_2011},
m: \citet{Shoshi_2024},
n: \citet{Flores_2023},
o: \citet{Natta_2006},
p: \citet{RuizRodriguez_2016},
q: \citet{Cieza_2010},
r: \citet{RuizRodriguez_2025},
s: \citet{Yen_2024},
t: \citet{McClure_2010},
u: \citet{Furlan_2009},
v: \citet{Flores_2022},
x: \citet{Shoshi_2025_oph},
y: \citet{Manara_2019},
z: \citet{Jiang_2026}.
}
\end{longtable}
\end{landscape}}
\twocolumn

\begin{figure*}[ht]
    \begin{center}
        \includegraphics[width=\linewidth]{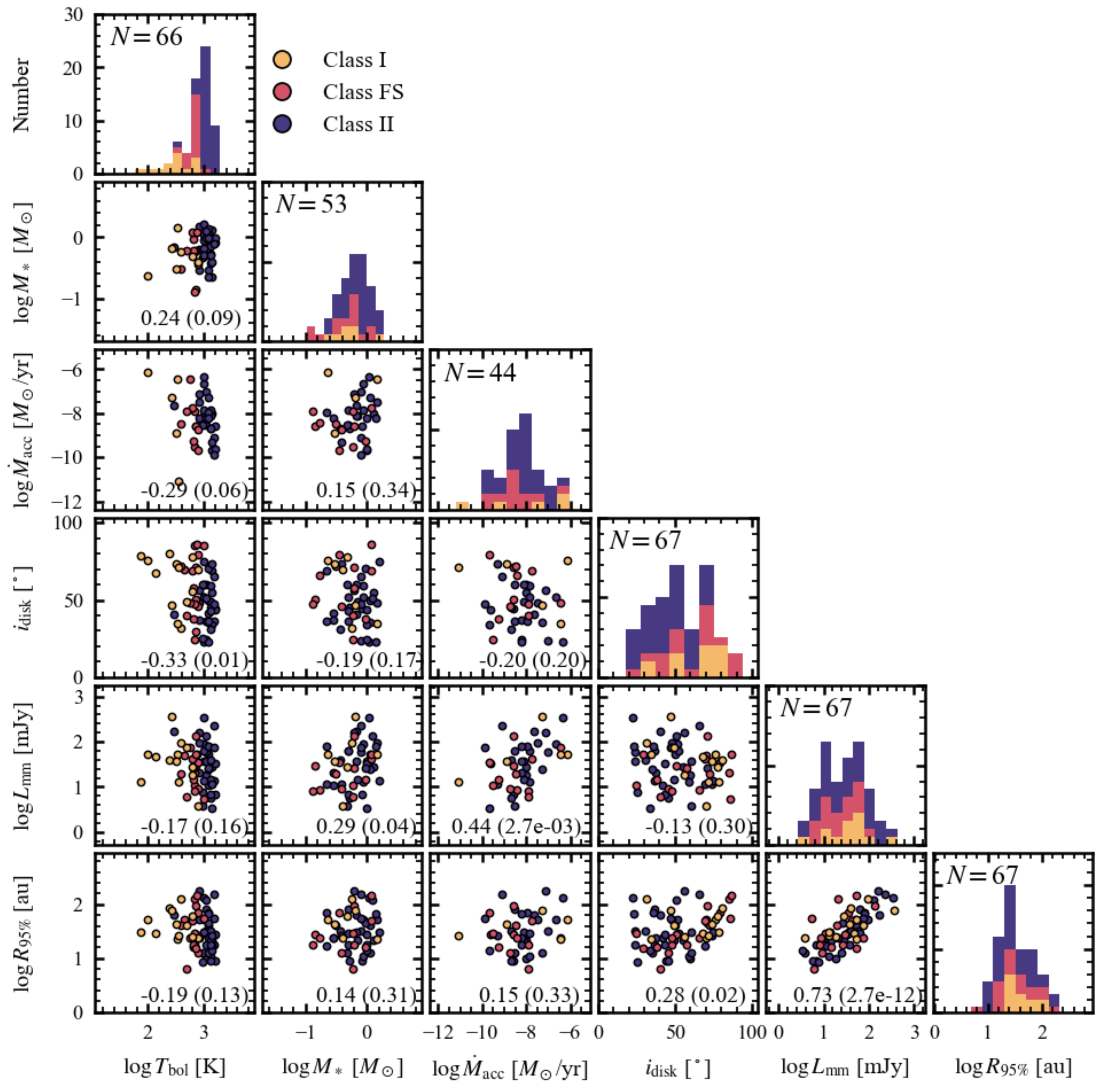}
    \end{center}
    \caption{
             Correlation matrix of six stellar and dust disk properties such as bolometric temperature $T_{\rm bol}$, stellar mass $M_\ast$, mass accretion rate $\dot{M}_{\rm acc}$, inclination angle $i_{\rm disk}$, millimeter luminosity $L_{\rm mm}$ (flux density scaled at the distance 140\,pc), and dust disk radius $R_{95\%}$.
             The top panels in each column show the histogram and the total number of samples for each parameter.
             The other panels show logarithmic scatter diagrams of the properties, except for the inclination angle $i_{\rm disk}$, with the Pearson correlation coefficients $\rho$ and $p$-values in brackets at the bottom.
             The yellow, red, and violet colors represent the disk evolutionary stages, namely Class~I, Class~FS, and Class~II, respectively.
             {Alt text: Correlation matrix of six stellar and disk properties. The diagonal panels show histograms, and the off-diagonal panels show scatter diagrams. 
             They summarize the sample size for each parameter and indicate whether a correlation is present for each combination.}
             }
    \label{fig:matrix}
\end{figure*}
\clearpage

\begin{table*}[ht]
    \caption{Correlation coefficients $\rho$ and $p$-values for each combination of the stellar and disk properties}
    \label{table:correlation}
    \centering
    \renewcommand{\arraystretch}{1.2}
    \begin{tabular}{cccccccc}
    \hline
    Combination & \multicolumn{2}{c}{All} & \multicolumn{2}{c}{Class~I and FS} & \multicolumn{2}{c}{Class~II} & Sample Size \\
                & $\rho$ & $p$-value & $\rho$ & $p$-value & $\rho$ & $p$-value & Total (I, FS, II) \\
    \hline
    $\log\,M_\ast$-$\log\,T_{\rm bol}$ & 0.24 & 0.09 & 0.10 & 0.68 & -0.01 & 0.95 & 52 (7, 12, 33) \\
    $\log\,\dot{M}_{\rm acc}$-$\log\,T_{\rm bol}$ & -0.29 & 0.06 & -0.45 & 0.08 & -0.48$^\ddagger$ & 0.01 & 43 (5, 11, 27) \\
    $i_{\rm disk}$-$\log\,T_{\rm bol}$ & -0.33$^\ddagger$ & 0.01 & -0.17 & 0.36 & 0.09 & 0.59 & 66 (13, 17, 36) \\
    $\log\,L_{\rm mm}$-$\log\,T_{\rm bol}$ & -0.17 & 0.16 & -0.26 & 0.17 & -0.36$^\ddagger$ & 0.03 & 66 (13, 17, 36) \\
    $\log\,R_{95\%}$-$\log\,T_{\rm bol}$ & -0.19 & 0.13 & -0.08 & 0.68 & -0.27 & 0.11 & 66 (13, 17, 36) \\
    $\log\,\dot{M}_{\rm acc}$-$\log\,M_\ast$ & 0.15 & 0.34 & 0.13 & 0.65 & 0.18 & 0.37 & 41 (4, 11, 26) \\
    $i_{\rm disk}$-$\log\,M_\ast$ & -0.19 & 0.17 & -0.16 & 0.51 & -0.05 & 0.80 & 53 (7, 13, 33) \\
    $\log\,L_{\rm mm}$-$\log\,M_\ast$ & 0.29$^\ddagger$ & 0.04 & 0.33 & 0.16 & 0.25 & 0.16 & 53 (7, 13, 33) \\
    $\log\,R_{95\%}$-$\log\,M_\ast$ & 0.14 & 0.31 & 0.19 & 0.43 & 0.15 & 0.40 & 53 (7, 13, 33) \\
    $i_{\rm disk}$-$\log\,\dot{M}_{\rm acc}$ & -0.20 & 0.20 & -0.16 & 0.54 & -0.22 & 0.26 & 44 (5, 12, 27) \\
    $\log\,L_{\rm mm}$-$\log\,\dot{M}_{\rm acc}$ & 0.44$^\dagger$ & 2.7$\times$10$^{-3}$ & 0.44 & 0.08 & 0.46$^\ddagger$ & 0.02 & 44 (5, 12, 27) \\
    $\log\,R_{95\%}$-$\log\,\dot{M}_{\rm acc}$ & 0.15 & 0.33 & 0.02 & 0.94 & 0.25 & 0.20 & 44 (5, 12, 27) \\
    $\log\,L_{\rm mm}$-$i_{\rm disk}$ & -0.13 & 0.30 & -0.09 & 0.65 & -0.15 & 0.37 & 67 (13, 18, 36) \\
    $\log\,R_{95\%}$-$i_{\rm disk}$ & -0.28$^\ddagger$ & 0.02 & 0.53$^\dagger$ & 2.0$\times$10$^{-3}$ & 0.01 & 0.97 & 67 (13, 18, 36) \\
    $\log\,R_{95\%}$-$\log\,L_{\rm mm}$ & 0.73$^\dagger$ & 2.7$\times$10$^{-12}$ & 0.53$^\dagger$ & 2.4$\times$10$^{-3}$ & 0.88$^\dagger$ & 1.7$\times$10$^{-12}$ & 67 (13, 18, 36) \\
    \hline
    \multicolumn{8}{p{0.9\linewidth}}{\footnotesize
    {\bf Note:} 
    A $\dagger$ symbol shows a statistically robust significant pair with the $p$-value$<0.003$, while a $\ddagger$ symbol represents a tentative pair with $0.003 \leq p < 0.05$.
    }\\
    \end{tabular}
    \label{tab:placeholder}
\end{table*}

\begin{figure*}[ht]
    \begin{center}
        \includegraphics[width=0.73\linewidth]{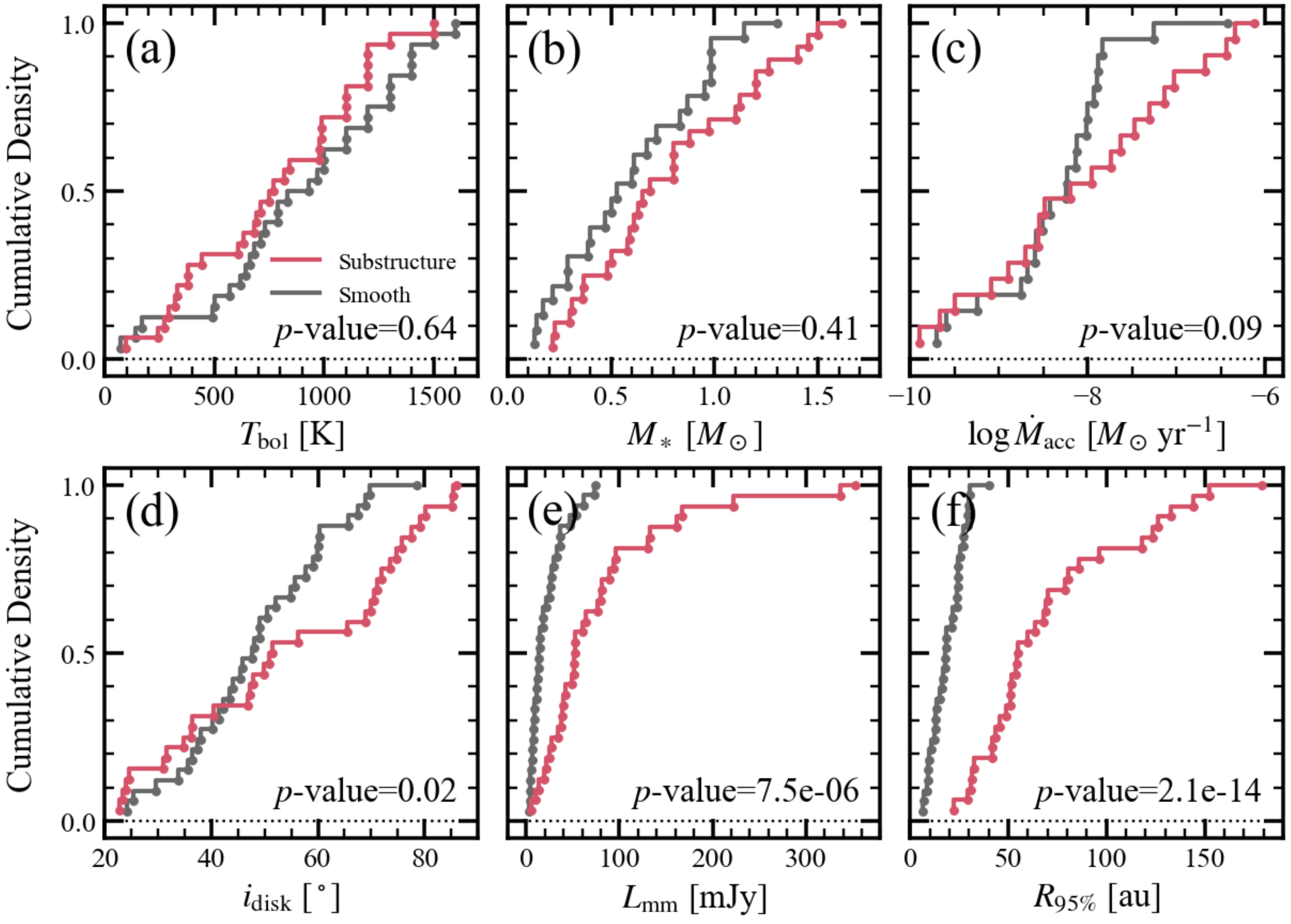}
    \end{center}
    \caption{
             Cumulative density functions of disks with substructures categorized as ``Ring'', ``Spiral'', and ``Inflection'' (red line) and smooth disks (gray line) based on (a) bolometric temperature $T_{\rm bol}$, (b) stellar mass $M_\ast$, (c) mass accretion rate $\dot{M}_{\rm acc}$, (d) inclination angle $i_{\rm disk}$, (e) millimeter luminosity $L_{\rm mm}$, and (f) dust disk radius $R_{95\%}$.
             The value at the lower right of each panel denotes the $p$-value derived by using the two-sample KS-test.
             Note that we used only the 67 Ophiuchus disks and did not include the eDisk sources used in Paper~\citetalias{Shoshi_2025_oph}. 
             {Alt text: The cumulative density functions indicate that the existence of disk substructures depends on $i_{\rm disk}$, $L_{\rm mm}$, and $R_{95\%}$.}
             }
    \label{fig:cdfsubst}
\end{figure*}

%%% Results %%%
\section{Statistical trends among stellar and disk properties}\label{sec:results}
Figure~\ref{fig:matrix} provides an overview of the correlation matrix among the six stellar and dust disk properties.
Table~\ref{table:correlation} summarizes the Pearson correlation coefficients and $p$-values for each parameter pair for three subsamples: all disks, Class~I/FS disks, and Class~II disks.
In this section, we first summarize the statistically significant correlations in section~\ref{subsec:corrsig} and then examine how the presence or absence of disk substructures relates to these stellar and disk properties in section~\ref{subsec:corrsub}.

\subsection{Statistically significant correlations and marginal trends}\label{subsec:corrsig}
The correlation matrix in figure~\ref{fig:matrix} and the statistics listed in table~\ref{table:correlation} show that only a subset of the parameter pairs exhibits clear departures from random scatter.
Adopting the significance thresholds based on the $p$-values, we identify one statistically robust correlation in the full sample, namely the $R_{95\%}$--$L_{\rm mm}$ relation, and four tentative correlations: $i_{\rm disk}$--$T_{\rm bol}$, $L_{\rm mm}$--$M_\ast$, $L_{\rm mm}$--$\dot{M}_{\rm acc}$, and $R_{95\%}$--$i_{\rm disk}$.
The $R_{95\%}$--$L_{\rm mm}$ relation is also found in both the Class~I/FS and Class~II subsamples, with moderate and strong strengths, respectively.
The $L_{\rm mm}$--$M_\ast$ and $L_{\rm mm}$--$\dot{M}_{\rm acc}$ relations show comparable strengths in the full sample and in the Class~I/FS and Class~II subsamples, but their statistical significance is either tentative or not significant.
The $R_{95\%}$--$i_{\rm disk}$ relation shows a robust positive correlation only in the Class~I/FS subsample.
This may partly reflect geometrical or inclination-dependent optical-depth effects, because highly inclined disks can appear larger if the emission remains optically thick out to larger radii.
However, because the same behavior is not seen in the full sample or in the Class~II subsample, we regard its physical interpretation as uncertain and do not discuss it further.
We instead focus on the other four parameter pairs, including the $i_{\rm disk}$--$T_{\rm bol}$ relation, which may be related to possible misclassification of evolutionary stages, and discuss their physical implications in detail in section~\ref{sec:discussion}.

By contrast, all other pairs have $p$-values greater than 0.05, indicating no statistically significant trends in our current sample.
In many cases, the Pearson coefficients satisfy $\lvert\rho\rvert\leq 0.3$, consistent with random scatter, especially when the sample is further divided into Class~I/FS and Class~II disks.
For several pairs involving Class~I/FS disks, fewer than 20 objects have measurements for both quantities, which limits the statistical power of our analysis.
We note that the $\dot{M}_{\rm acc}$-$T_{\rm bol}$ pair yields weak-to-moderate correlation coefficients ($\lvert\rho\rvert\sim$0.3-0.5) for the full sample and for both subsamples.
However, the corresponding $p$-values are not consistent across subsamples: for Class~II disks, the $p$-value$=0.01$, whereas for the full sample and the Class~I/FS subsample, the $p$-value=0.06 and the $p$-value=0.08, respectively.
Given this lack of consistency and the limited sample sizes, we treat the $\dot{M}_{\rm acc}$-$T_{\rm bol}$ relation as a possible but statistically insignificant trend that should be revisited with larger and more homogeneous samples.
Such samples, spanning a wider range of evolutionary stages and including more Class~I and FS systems, will be required to robustly assess these correlations and distinguish genuine physical relations from artifacts of small-number statistics.

\subsection{Disk substructure dependence}\label{subsec:corrsub}
In Paper~\citetalias{Shoshi_2025_oph}, we categorized each disk as ``Ring'', ``Spiral'', ``Inflection'', or ``Smooth'' based on the intensity profile derived from the PRIISM images.
By combining our targets with the eDisk sample, we previously identified that the appearance of detectable substructures depends on the bolometric temperature $T_{\rm bol}$ and the radius of the dust disk $R_{95\%}$.
In this work, we use the additional stellar and disk parameters compiled in table~\ref{table:properties} to extend the comparison beyond $T_{\rm bol}$ and $R_{95\%}$.
We treat disks categorized as ``Ring'', ``Spiral'', and ``Inflection'' as \emph{substructured} disks and those classified as ``Smooth'' as \emph{smooth} disks.
To assess differences in the distribution of each parameter between these two groups in a conservative and non-parametric manner, we apply the Kolmogorov-Smirnov two-sample test \citep[KS test;] []{Wall_2012} to our sample.

Figure~\ref{fig:cdfsubst} shows the cumulative distribution functions of disks with ``smooth'' distributions and ``substructures'' as a function of $T_{\rm bol}$, $M_\ast$, $\dot{M}_{\rm acc}$, $i_{\rm disk}$, $L_{\rm mm}$, and $R_{95\%}$, together with the corresponding KS-test $p$-values in each panel.
The associated scatter plots are presented in appendix~\ref{sec:other_scatter}.
For the stellar parameters ($T_{\rm bol}$, $M_\ast$, and $\dot{M}_{\rm acc}$), we do not find statistically significant differences between the substructured and smooth populations within the Ophiuchus-only sample (figure~\ref{fig:cdfsubst}a-c).
We note that this Ophiuchus-only sample lacks smooth disks at $100\,{\rm K}\leq T_{\rm bol}\leq 500\,{\rm K}$, which limits our ability to test the $T_{\rm bol}$ dependence inferred from the combined sample in Paper~\citetalias{Shoshi_2025_oph}.
The absence of a clear separation in $M_\ast$ suggests that, within the low--mass regime probed by our sample, the occurrence of detectable substructures could not appear to be primarily controlled by stellar mass.
Furthermore, the KS test for $\dot{M}_{\rm acc}$ between disks with substructures and smooth disks gives the $p$-value=0.09, which does not satisfy our significance criterion, where we do not find statistically significant evidence that the occurrence of detectable substructures depends on $\dot{M}_{\rm acc}$ in the present sample.

In contrast to the stellar properties, the disk properties show statistically significant differences between substructured and smooth disks.
The inclination angle $i_{\rm disk}$ (figure~\ref{fig:cdfsubst}d) shows a tentative difference between the two populations (the $p$-value=0.02). 
Since $i_{\rm disk}$ is a viewing-angle parameter and substructures such as rings and gaps are generally more difficult to identify in highly inclined disks, we do not interpret this tentative difference as evidence for a physical connection between inclination and substructure formation.
The largest difference is found for the millimeter luminosity and dust radius (figure~\ref{fig:cdfsubst}e and \ref{fig:cdfsubst}f).
For $L_{\rm mm}$, the KS test yields the $p$-value=$1.4\times10^{-5}$, indicating that smooth disks are concentrated at lower luminosities. 
In contrast, disks with substructures are more prevalent toward the bright end of the distribution.
An even stronger separation is seen for $R_{95\%}$ (the $p$-value=$4.5\times10^{-14}$), where substructured disks systematically extend to larger radii than smooth ones.
Given the strong $R_{95\%}$-$L_{\rm mm}$ correlation and the fact that substructure detectability is primarily bounded in $R_{95\%}$ (Paper~\citetalias{Shoshi_2025_oph}), the separation in $L_{\rm mm}$ is likely at least partly driven by the separation in $R_{95\%}$.
Overall, these trends are consistent with detectable substructures being more common in radially extended, dust-rich disks.
The trends may partly reflect a detection bias, since larger disks are sampled by more resolution elements, but they may also reflect an intrinsic difference between extended disks and compact, faint disks.
Indeed, \citet{Bhowmik_2026} found from ALMA Band~8 observations, with resolutions of $\sim$7-20~au, that substructures are detected less frequently in compact Ophiuchus disks than in larger disks.
Similarly, recent high-resolution surveys in other regions suggest that substructures are more commonly detected in extended disks than in compact disks \citep[][]{Bosschaart_2026,Guerra-Alvarado_2025}.
In particular, \citet{Guerra-Alvarado_2025} suggested that compact disks may be drift-dominated, possibly lacking deep dust traps that halt radial drift.
Determining whether the small disks indeed lack substructures will require deeper, higher-resolution observations that can sample compact, faint disks with a number of resolution elements comparable to those of large, bright disks.

In summary, our Ophiuchus-only available analysis indicates that the presence of detectable millimeter substructures is most strongly related to the dust disk properties $L_{\rm mm}$ and $R_{95\%}$. 
We do not find statistically significant evidence for additional dependencies on the other stellar parameters tested here within the current sample. 
Note that our sample is defined by the resolution and sensitivity criteria described in section~\ref{subsec:selection}, and is therefore less complete for compact and/or faint disks (approximately $L_{\rm mm}\leq 20$\,mJy and $R_{95\%}\leq 20$\,au), which may cause the cumulative distribution functions to under-represent the small and faint end. 
Moreover, for the five targets (ISO-Oph~59, DoAr~32, ISO-Oph~155, ISO-Oph~20, and ISO-Oph~116), the baseline lengths required to detect disk substructures using the PRIISM imaging are predicted to exceed the maximum baseline length of the observations ($\sim $1\,M$\lambda$), implying limited sensitivity to substructures for these sources (appendix~2.2 of Paper~\citetalias{Shoshi_2025_oph}). 
A larger, more homogeneous dataset with higher sensitivity and maximum baselines of $\gtrsim$2\,M$\lambda$ will be essential to test whether further dependencies emerge beyond those reported here.

\begin{figure}[t]
    \begin{center}
        \includegraphics[width=\linewidth]{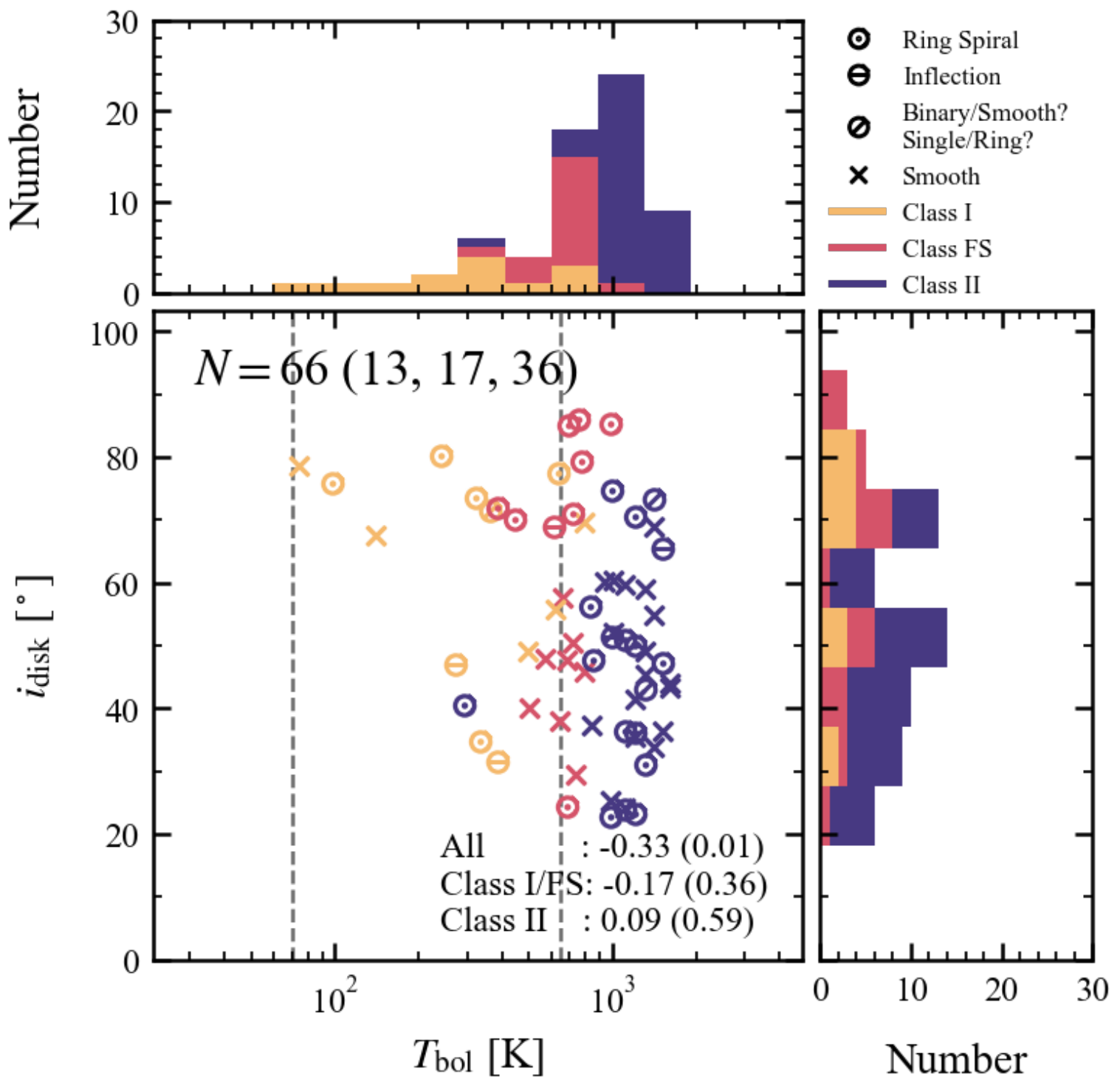}
    \end{center}
    \caption{
             Relationship between bolometric temperature $T_{\rm bol}$ and dust disk inclination angle $i_{\rm disk}$ for 66 disks.
             The colors of the symbols indicate the evolutionary stage classified by the spectral slope at 2-22\,$\mu$m, where yellow, red, and violet correspond to Class~I, FS, and Class~II stages.
             The marker shapes denote the disk categorizations in Paper~\citetalias{Shoshi_2025_oph}; $\odot$ as ``Ring'' or ``Spiral'', $\ominus$ as ``Inflection'', $\oslash$ as candidates for nearly edge-on disks with ``Ring'' features or circumstellar disks in binary systems, and $\times$  as ``Smooth'' brightness distributions.
             The values at the lower right of the scatter diagram show the Pearson correlation coefficients $\rho$, with the corresponding $p$-values in brackets, for all disks, Class~I and FS disks, and Class~II disks.
             The gray dashed lines indicate the thresholds of 70\,K between Class~0 and Class~I stages, and 650\,K between Class~I and Class~II stages.
             The top panel shows the histogram of $T_{\rm bol}$, and the right panel shows that of $i_{\rm disk}$, indicating the sample size for each quantity.
             {Alt text: The scatter diagram of $T_{\rm bol}$-$i_{\rm disk}$ shows that the main locus of sources transitions smoothly from Class~I through FS to Class~II around $T_{\rm bol}= 650$\,K, indicating that $i_{\rm disk}$ could not affect the classification based on $T_{\rm bol}$.}
             }
    \label{fig:tbol_inc}
\end{figure}

\begin{figure*}[t]
    \begin{center}
        \includegraphics[width=\linewidth]{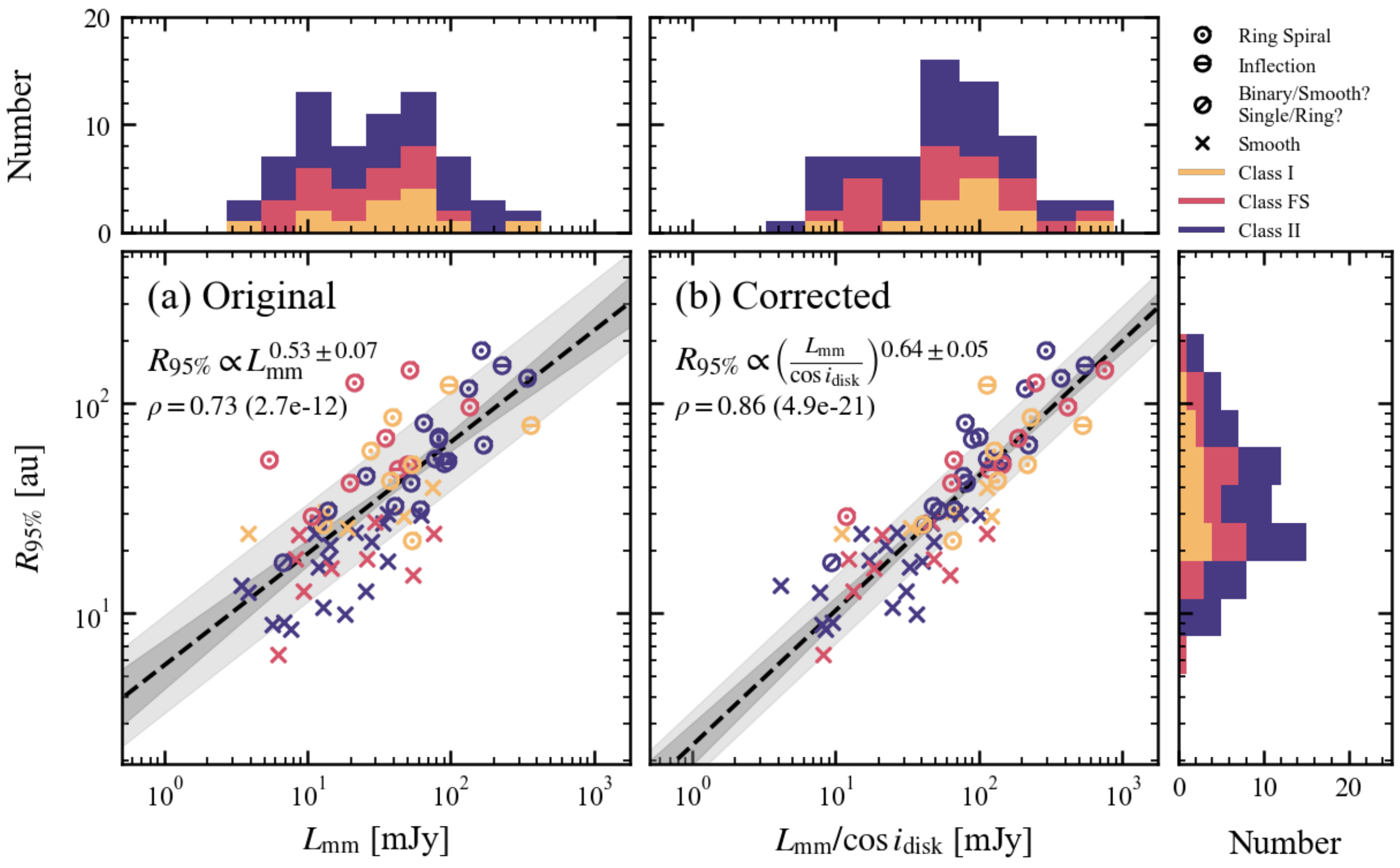}
    \end{center}
    \caption{
             Relationship of dust disk radius $R_{95\%}$ with (a) the original millimeter luminosity $L_{\rm mm}$ and (b) the millimeter luminosity corrected by the inclination angle $L_{\rm mm}/\cos\,i_{\rm disk}$ for 67 disks.
             The marker shapes denote the disk categorizations; $\odot$ as ``Ring'' or ``Spiral'', $\ominus$ as ``Inflection'', $\oslash$ as candidates for nearly edge-on disks with ``Ring'' features or circumstellar disks in binary systems, and $\times$  as ``Smooth'' brightness distributions.
             The colors of the symbols indicate the evolutionary stage classified by the spectral slope at 2-22\,$\mu$m, where yellow, red, and violet correspond to Class~I, FS, and Class~II stages.
             In each panel, the black dashed line indicates the median scaling relation obtained from Bayesian linear regression, and the dark gray shaded region represents the 68\% confidence interval around it. 
             The light gray shaded region corresponds to the intrinsic scatter inferred by the regression. 
             The best-fitting linear regression parameters and Pearson correlation coefficients $\rho$ are listed in the upper left corner of each panel, with the associated $p$-values shown in brackets.
             The top panel shows the histogram of $L_{\rm mm}$ or $L_{\rm mm}/\cos\,i_{\rm disk}$, and the right panel shows that of $R_{95\%}$, indicating the sample size for each quantity.
             {Alt text: The scatter diagrams of dust disk radius $R_{95\%}$ with the original millimeter luminosity $L_{\rm mm}$ and the luminosity corrected by the inclination angle $L_{\rm mm}/\cos\,i_{\rm disk}$ represent the strong correlations.}
             }
    \label{fig:lmm_radi}
\end{figure*}

\begin{figure*}[t]
    \begin{center}
        \includegraphics[width=\linewidth]{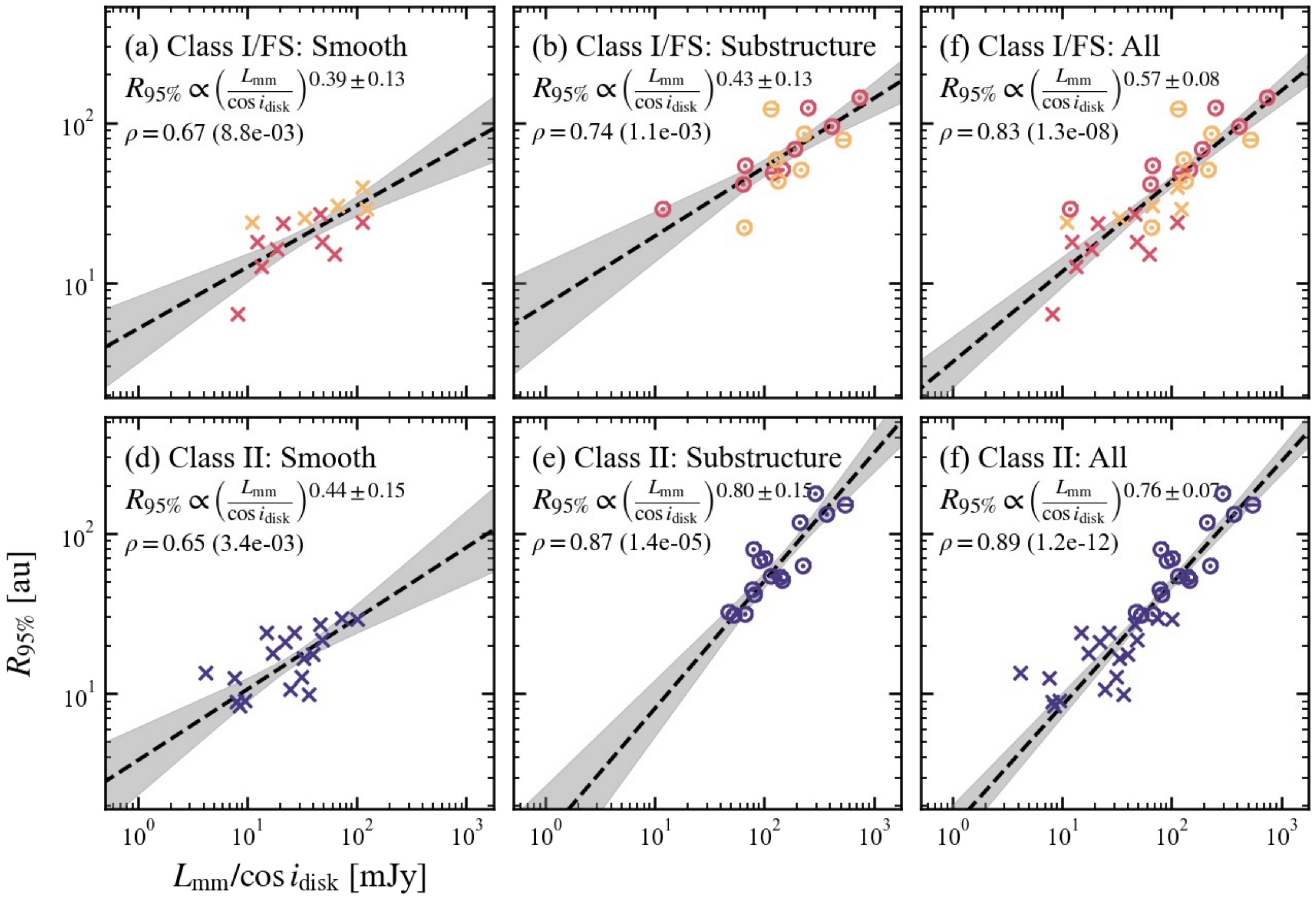}
    \end{center}
    \caption{
             Relationship between the millimeter luminosity corrected by the inclination angle $L_{\rm mm}/\cos i_{\rm disk}$ and the dust radius $R_{95\%}$ for each evolutionary stage and the presence of substructures.
             The top panels show Class~I and FS disks, and the bottom panels show Class~II disks. 
             Left, middle, and right columns correspond to disks with smooth brightness distribution (``Smooth''), any substructure (``Ring'', ``Spiral", and ``Inflection''), and all the disks.
             The marker shapes denote the disk categorizations; $\odot$ as ``Ring'' or ``Spiral'', $\ominus$ as ``Inflection'', $\oslash$ as candidates for nearly edge-on disks with ``Ring'' features or circumstellar disks in binary systems, and $\times$  as ``Smooth'' brightness distributions.
             The colors of the symbols indicate the evolutionary stage classified by the spectral slope at 2-22\,$\mu$m, where yellow, red, and violet correspond to Class~I, FS, and Class~II stages.
             In each panel, the black dashed line indicates the median scaling relation obtained from Bayesian linear regression, and the dark gray shaded region represents the 68\% confidence interval around it. 
             %The light gray shaded region corresponds to the intrinsic scatter inferred by the regression. 
             The best-fitting linear regression parameters and Pearson correlation coefficients $\rho$ are listed in the upper left corner of each panel, with the associated $p$-values shown in brackets.
             {Alt text: The scatter diagrams between the millimeter luminosity corrected by inclination angle $L_{\rm mm}/\cos i_{\rm disk}$ and dust disk radius $R_{95\%}$ indicate that the relation differs between Class~I/FS and Class~II disks, being especially pronounced for Class~II disks with substructures.}
             }
    \label{fig:lmm_radi_class}
\end{figure*}

%%% Discussion %%%
\section{Discussion}\label{sec:discussion}
\subsection{Evaluation for SED classification}\label{subsec:tbol_inc}
The evolutionary stages of YSOs are usually classified based on their spectral energy distributions (SEDs), with the near- to mid-infrared spectral slope and the bolometric temperature $T_{\rm bol}$ commonly used as indicators \citep[e.g.,][]{Greene_1994,Chen_1995,Evans_2009}.
The spectral slope is evaluated between two wavelengths, whereas $T_{\rm bol}$ is computed from the full SED as
\begin{align}
    T_{\rm bol}=1.25\times10^{-11}\frac{\int^\infty_0\nu F_\nu\,d\nu}{\int^\infty_0F_\nu\,d\nu}\,[{\rm K}],\label{eq:tbol}
\end{align}
where $\nu$ and $F_\nu$ are the observed frequency and the flux density.
Radiative transfer calculations have shown that increasing inclination enhances extinction and absorption along the line of sight, suppressing the near-infrared emission and allowing evolved systems viewed nearly edge-on to appear younger in SED-based schemes \citep[e.g.,][]{Masunaga_2000,Furlan_2016}.
To address this point, we compared the classifications based on the spectral slope between 2\,$\mu$m and 22\,$\mu$m and those based on $T_{\rm bol}$. 
We found them to be broadly consistent, with no major systematic discrepancies across our targets (see appendix~4 of Paper~\citetalias{Shoshi_2025_oph}).
Given this consistency, we next examine the relationship between $T_{\rm bol}$ and the disk inclination $i_{\rm disk}$ to assess whether the effects caused by $i_{\rm disk}$ could still influence the inferred evolutionary stage within our sample.

Figure~\ref{fig:tbol_inc} represents the distribution of $T_{\rm bol}$ and $i_{\rm disk}$, with symbol colors indicating the SED classes derived from the 2-22\,$\mu$m spectral slope (Class~I, FS, and II; see section~\ref{subsec:selection}).
We find a weak but statistically tentative negative correlation in $T_{\rm bol}$ and $i_{\rm disk}$ when all disks are considered together ($\rho=-0.29$, the $p$-value=0.02), indicating that nearly edge-on systems tend to populate the low-$T_{\rm bol}$ range.
However, when the sample is divided into Class~I/FS and Class~II disks, no statistically significant or tentative correlation is seen in either subsample; in fact, the Class~II disks even show a weak positive coefficient.
The lack of a consistent trend across subsamples suggests that the global $i_{\rm disk}$-$T_{\rm bol}$ relation in the combined sample does not simply trace a universal physical dependence of $T_{\rm bol}$ on $i_{\rm disk}$.

Moreover, the main locus of sources transitions smoothly from Class~I through FS to Class~II around $T_{\rm bol}\approx 650$\,K, supporting the overall consistency between the classifications based on the 2-22\,$\mu$m spectral slope and those based on $T_{\rm bol}$ for the bulk of our sample.
As a caveat, although ISO-Oph~17 is a Class~II source with an estimated age of 1-2\,Myr \citep[][]{Andrews_2018_dsharp,Esplin_2020}, it exhibits a relatively low bolometric temperature ($T_{\rm bol}=270$\,K), possibly because it is located in a high-extinction region of the L1688 cluster.

The apparent anti-correlation between $T_{\rm bol}$ and $i_{\rm disk}$ in the combined sample can instead be understood as a consequence of sample selection and incompleteness in inclination space.
As discussed in Paper~\citetalias{Shoshi_2025_oph}, Class~II disks in our sample exhibit a deficit of nearly edge-on systems relative to a random $\cos i_{\rm disk}$ distribution.
In contrast, Class~I sources with $T_{\rm bol}\sim$100-300\,K are sparsely sampled, particularly nearly face-on disks.
When these effects are combined, they naturally produce a pattern in which low-$T_{\rm bol}$ bins are preferentially populated by highly inclined objects and higher-$T_{\rm bol}$ bins by more moderately inclined disks, even if $T_{\rm bol}$ itself is not strongly controlled by geometry.

Overall, we therefore conclude that, within the limitations of the present sample, $i_{\rm disk}$ does not appear to introduce a dominant systematic bias in the $T_{\rm bol}$-based classification.
In other words, because $T_{\rm bol}$ is defined from the full SED and is therefore less sensitive to viewing geometry than the spectral slope at two wavelengths, we regard the bolometric temperature $T_{\rm bol}$ as a relatively robust and convenient statistical tracer of evolutionary stage.
Larger and more homogeneous samples, particularly including more Class~I and FS systems spanning $T_{\rm bol}\sim100$-300\,K, will be essential to test and refine these conclusions.

\begin{table*}[ht]
\caption{Results for the Bayesian linear regression for the relation between corrected millimeter luminosity $L_{\rm mm}/\cos\,i_{\rm disk}$ and dust disk radius $R_{95\%}$.}
\label{table:lmm_radii}
\begin{center}
\renewcommand{\arraystretch}{1.2}
\begin{tabular}{ccccccc}
\hline
Evolutionary Stage & Disk Categorization & Sample Size & Intercept & Slope   & Dispersion        & Panel of Figure~\ref{fig:lmm_radi_class}\\
                   &                     &             & $\alpha$  & $\beta$ & $\sigma_{\rm SD}$ & \\
                   &                     &             &           &         & dex               & \\
\hline
               & Smooth       & 14 & $0.71\pm0.21$ & $0.39\pm0.13$ & $0.21$ & (a) \\
Class~I and FS & Substructure & 16 & $0.86\pm0.28$ & $0.43\pm0.13$ & $0.13$ & (b) \\
               & All          & 30 & $0.50\pm0.15$ & $0.57\pm0.08$ & $0.15$ & (c) \\
         & Smooth       & 18 & $0.58\pm0.21$ & $0.44\pm0.15$ & $0.17$ & (d) \\
Class~II & Substructure & 16 & $0.11\pm0.33$ & $0.80\pm0.15$ & $0.17$ & (e) \\
         & All          & 34 & $0.16\pm0.13$ & $0.76\pm0.07$ & $0.19$ & (f) \\
    & Smooth       & 32 & $0.61\pm0.13$ & $0.44\pm0.08$ & $0.19$ & -- \\
All & Substructure & 32 & $0.63\pm0.20$ & $0.55\pm0.09$ & $0.14$ & -- \\
    & All          & 64 & $0.34\pm0.09$ & $0.65\pm0.05$ & $0.18$ & -- \\
\hline
\multicolumn{7}{p{0.9\linewidth}}{\footnotesize
{\bf Note:} 
In this analysis, we treated the categorizations of ``Ring'', ``Spiral'', and ``Inflection'' as substructures and remove candidates for nearly edge-on disks with ``Ring'' features or circumstellar disks in binary systems.
}\\
\end{tabular}
\end{center}
\end{table*}

\subsection{Evolutionary stage and substructure dependence in the $R_{95\%}$-$L_{\rm mm}$ relation}\label{subsec:lmm_r95}
The correlation between millimeter luminosity and dust disk radius in nearby star-forming regions has been investigated using Submillimeter Array (SMA) or ALMA observations \citep[e.g.,][]{Tripathi_2017,Tazzari_2017,Barenfeld_2017,Andrews_2018_sma,Long_2019,Hendler_2020}.
In previous studies, disk radii have typically been inferred either from parametric modeling of the radial emission profile in the image plane or from fitting parametric models directly in the visibility domain.
We instead measure disk radii in the image domain from PRIISM super-resolution images in this work, which provide a homogeneous and largely model-independent characterization of the two-dimensional brightness distribution for 67 spatially resolved disks \citep[][]{Yamaguchi_2024,Shoshi_2026}.
In the following, we examine the relation between the millimeter luminosity $L_{\rm mm}$ and the dust disk radius $R_{95\%}$ for our Ophiuchus sample.

Figure~\ref{fig:lmm_radi}(a) shows the millimeter luminosity $L_{\rm mm}$ as a function of dust disk radius $R_{95\%}$ for all the Ophiuchus disks.
As described in section~\ref{subsec:corrsig}, we see the strongest correlation with $\rho=0.73$ and the $p$-value of $2.7\times10^{-4}$.
To quantify this correlation, we perform Bayesian linear regression in logarithmic space using the Python module \texttt{Linmix} \citep{Kelly_2007} and yield
\begin{align}
    \log\,\left(\frac{R_{95\%}}{{\rm au}}\right)=\left(0.73^{+0.10}_{-0.10}\right)+\left(0.54^{+0.06}_{-0.06}\right)\log\,\left(\frac{L_{\rm mm}}{1\,{\rm mJy}}\right),\label{eq:lmm_all}
\end{align}
with an intrinsic scatter of 0.23\,dex.
The value of the slope is roughly comparable to the estimates in \citet{Hendler_2020} and \citet{Dasgupta_2025} and is consistent with the model of dust radial drift \citep[][]{Rosotti_2019}.

To approximately account for the optical depth effects caused by the inclination angle $i_{\rm disk}$, which was discussed by \citet{Tazzari_2021}, we reanalyze the correlation between millimeter luminosity and $R_{95\%}$ using a luminosity rescaled by $\cos\,i_{\rm disk}$ (i.e., $L_{\rm mm}/\cos\,i_{\rm disk}$).
Figure~\ref{fig:lmm_radi}(b) shows the relationship between the corrected millimeter luminosity $L_{\rm mm}/\cos\,i_{\rm disk}$ and $R_{95\%}$.
We employ the Bayesian linear regression using \texttt{Linmix} to obtain the relation represented as
\begin{align}
    \log\,\left(\frac{R_{95\%}}{{\rm au}}\right)=\left(0.37^{+0.09}_{-0.09}\right)+\left(0.64^{+0.05}_{-0.05}\right)\log\,\left(\frac{L_{\rm mm}/\cos\,i_{\rm disk}}{1\,{\rm mJy}}\right),\label{eq:corrlmm_all}
\end{align}
with an intrinsic dispersion of 0.17\,dex.
We confirm a smaller scatter in the correlation when the $\cos\,i_{\rm disk}$ effect is considered, where the correlation coefficient $\rho$ is improved by $\sim$18\% ($\rho$=0.86 with the $p$-value of $4.9\times10^{-21}$) compared to the case without the consideration.
This improvement likely reflects that our PRIISM images spatially resolve the disks and thus enable more robust, object-by-object inclination estimates, making the $\cos i_{\rm disk}$ correction more effective.

We confirm statistically significant $R_{95\%}$-$L_{\rm mm}$ correlations in both the Class~I/FS and Class~II subsamples, and find that disks with substructures preferentially occupy the large-$L_{\rm mm}$ and large-$R_{95\%}$ regime in our Ophiuchus sample (see section~\ref{subsec:corrsub}).
Motivated by these trends, we examine how evolutionary stage and the presence of substructures affect the $R_{95\%}$-$L_{\rm mm}/\cos i_{\rm disk}$ relation.
Figure~\ref{fig:lmm_radi_class} shows the $L_{\rm mm}/\cos i_{\rm disk}$-$R_{95\%}$ distribution separated by evolutionary stage (Class~I/FS and Class~II) and morphology (smooth disks, disks with substructures, and all disks).
We classify sources labeled as ``Ring'', ``Spiral'', and ``Inflection'' as disks with substructures.
We exclude sources classified as ``Single/Ring or Binary/Smooth'' in Paper~\citetalias{Shoshi_2025_oph} since it is unclear whether the sources are nearly edge--on disks with rings or circumstellar disks in binary systems.
In both the Class~I/FS and Class~II subsamples, disks with substructures tend to populate the higher-$L_{\rm mm}/\cos i_{\rm disk}$ and larger-$R_{95\%}$ end of the distribution.

To quantify these relations, we fit each subsample with the form
\begin{align}
    \log\left(\frac{R_{95\%}}{{\rm au}}\right) = \alpha+ \beta\log\left(\frac{L_{\rm mm}/\cos i_{\rm disk}}{1\,{\rm mJy}}\right),\label{eq:corrlmm_sample}
\end{align}
including an intrinsic scatter $\sigma_{\rm SD}$.
We perform Bayesian linear regression using \texttt{Linmix} for each subsample in the $R_{95\%}$-$L_{\rm mm}/\cos i_{\rm disk}$ plane, and summarize the inferred values of the intercept $\alpha$, the slope $\beta$, and the dispersion $\sigma_{\rm SD}$ in table~\ref{table:lmm_radii}.

The inferred slopes indicate that $\beta$ depends on the evolutionary stage and the presence of disk substructures.
In the Class~I/FS stage, disks with and without substructures have comparable slopes, with $\beta=0.43\pm0.13$ and $\beta=0.39\pm0.13$, respectively (figure~\ref{fig:lmm_radi_class}a,b).
On the other hand, in the Class~II stage, the dependence of the substructure is more pronounced.
Smooth disks have $\beta=0.44\pm0.15$, while substructured disks have $\beta=0.80\pm0.15$ (figure~\ref{fig:lmm_radi_class}d,e), indicating a departure from the scaling $\beta\sim0.4$ seen in the Class~I/FS sample.
As a result, the steep slope of the Class~II disks with substructures increases the slope measured for the full Class~II sample ($\beta=0.76\pm0.07$), yielding a steeper trend than that for the full Class~I/FS sample ($\beta=0.57\pm0.08$).

The comparison of the $R_{95\%}$-$L_{\rm mm}/\cos i_{\rm disk}$ relation across evolutionary stages and the presence of substructures shows that only Class~II disks with substructures follow a distinct trend, with $\beta=0.80\pm0.15$.
Clarifying which physical processes drive this difference is key to interpreting our results.
In this context, \citet{Zormpas_2022} reported a systematic exploration of the temporal evolution of continuum emission and disk size using one-dimensional viscous evolution models that include dust growth, fragmentation, and radial drift.
Their model suite considers both smooth disks and disks with planet-induced gap structures (pressure bumps), exploring planet-to-star mass ratios of $q=3\times10^{-4}$ and $10^{-3}$, corresponding to roughly sub-Jovian to Jovian-mass companions for a solar-mass host star.
In their framework, these pressure bumps correspond to pressure maxima associated with planet-opened gaps that are sufficiently strong to trap dust and affect the global size-luminosity relation.
Note that the 68\% flux radius $r_{\rm eff}$ adopted by \citet{Zormpas_2022} differs from our definition of $R_{95\%}$, but in our sample $R_{95\%}$ and $R_{68\%}$ are approximately linearly related (see appendix~\ref{sec:r68r95}), so that this difference should have only a minor impact on comparisons of power-law indices.
\citet{Zormpas_2022} found that smooth disks can follow the observationally established scaling $L_{\rm mm}\propto r_{\rm eff}^{2}$ (i.e., $r_{\rm eff}\propto L_{\rm mm}^{0.5}$). 
On the other hand, disks with strong dust traps can evolve toward $L_{\rm mm}\propto r_{\rm eff}^{5/4}$ (i.e., $r_{\rm eff}\propto L_{\rm mm}^{0.8}$) after $\gtrsim 1$\,Myr.
They further argued that a mixed population of smooth and strongly substructured disks can account for the observed spread in the $r_{\rm eff}$-$L_{\rm mm}$ diagram, with the large and bright region predominantly populated by disks with substructures and the small and faint region primarily occupied by smooth disks.
\citet{Zormpas_2022} modeled these traps as planet-induced gap structures, suggesting that planet-induced pressure maxima can regulate dust evolution in disks. 
This interpretation is consistent with the recent modeling by \citet{Orcajo_2025}, who showed that planet-induced substructures can reproduce the morphologies of Ophiuchus disks, as well as those of bright disks more generally.

These model results provide a useful framework for interpreting our data.
In particular, two aspects of our observed trends are consistent with the predictions of \citet{Zormpas_2022}.
First, only the Class~II disks with substructures exhibit a steep $R_{95\%}$-$L_{\rm mm}/\cos i_{\rm disk}$ scaling with $\beta\simeq0.8$. 
In contrast, the other subsamples remain broadly consistent with the shallower scaling expected for smooth disks (and for earlier stages with substructures), $R\propto L_{\rm mm}^{0.5}$.
Second, disks with substructures preferentially populate the large and bright end of the $R_{95\%}$-$L_{\rm mm}/\cos i_{\rm disk}$ plane. 
Conversely, smooth disks are more common toward the faint, small end, mirroring the mixed-population result proposed by \citet{Zormpas_2022}.
Taken together, these results are consistent with a scenario in which substructures may already form by the mid Class~I stage (Paper~\citetalias{Shoshi_2025_oph}) and subsequently promote efficient dust trapping that persists into the Class~II stage, leaving an imprint on the global size-luminosity scaling.

Nevertheless, several caveats should be kept in mind when interpreting our results.
As discussed in Paper~\citetalias{Shoshi_2025_oph}, limitations in the maximum baseline length and the difficulty of identifying substructures in compact disks with radii $\lesssim$15-20\,au can influence the inferred relations, mainly by making the substructure census less uniform than the size measurements.
In this sense, the present data still provide useful constraints on global disk sizes (and thus on the $R_{95\%}$-$L_{\rm mm}$ relation).
At the same time, a more complete assessment of substructure occurrence across the full range of disk sizes will benefit from additional angular resolution and sensitivity.
In the context of dust-trap evolution models, the morphology-dependent $R_{95\%}$-$L_{\rm mm}/\cos i_{\rm disk}$ scaling found here further suggests that disks classified as ``smooth'' at the current resolution may not constitute a single physical population.
Some systems may genuinely lack strong, long-lived pressure bumps (e.g., because gap-opening planets have not yet formed). 
In contrast, others may host substructures on smaller spatial scales and/or weaker dust traps that do not appreciably modify the global size-luminosity scaling.

Future high-resolution and high-sensitivity surveys, especially targeting the smooth and compact population, will therefore provide a valuable opportunity to distinguish between these possibilities and to map where, within the Class~II population, the $R_{95\%}$-$L_{\rm mm}/\cos i_{\rm disk}$ relations for smooth disks and disks with substructures begin to diverge and/or overlap.
If the steeper branch is indeed linked to dust trapping in pressure bumps, then corresponding structures should also be present in the gas distribution; spatially resolved molecular-line observations will therefore be essential for testing whether gas gaps or pressure maxima accompany the continuum substructures.

\begin{figure*}[ht]
    \begin{center}
        \includegraphics[width=\linewidth]{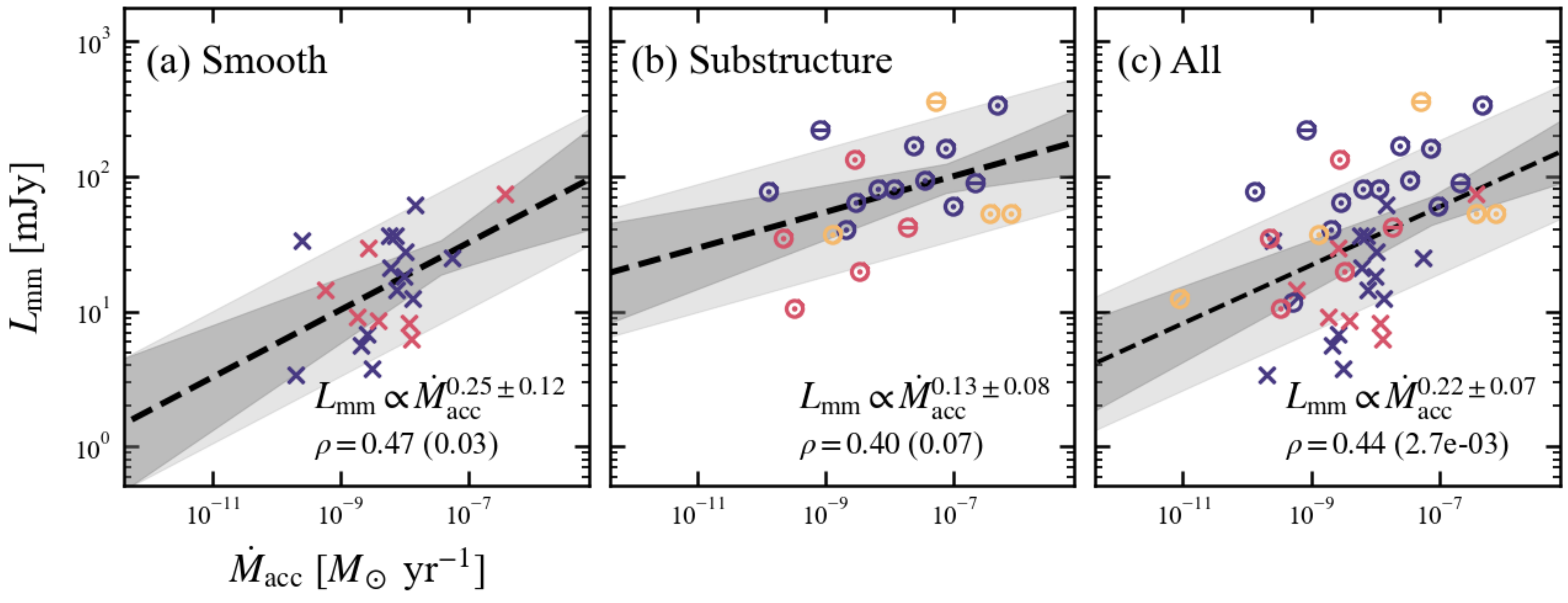}
    \end{center}
    \caption{
             Relationship between mass accretion rate $\dot{M}_{\rm acc}$ and millimeter luminosity $L_{\rm mm}$ for (a) 21 disks categorized as ``Smooth'', (b) 21 disks with substructures (``Ring'', ``Spiral'', and ``Inflection''), and (c) all 44 the disks, including ``Candidates''.
             The marker shapes denote the disk categorizations; $\odot$ as ``Ring'' or ``Spiral'', $\ominus$ as ``Inflection'', $\oslash$ as candidates for nearly edge-on disks with ``Ring'' features or circumstellar disks in binary systems, and $\times$  as ``Smooth'' brightness distributions.
             The colors of the symbols indicate the evolutionary stage classified by the spectral slope at 2-22\,$\mu$m, where yellow, red, and violet correspond to Class~I, FS, and Class~II stages.
             In each panel, the black dashed line indicates the median scaling relation obtained from Bayesian linear regression, and the dark gray shaded region represents the 68\% confidence interval around it. 
             The light gray shaded region corresponds to the intrinsic scatter inferred by the regression. 
             The best-fitting linear regression parameters and Pearson correlation coefficients $\rho$ are listed in the lower right corner of each panel, with the associated $p$-values shown in brackets.
             {Alt text: The scatter diagrams between the mass accretion rate $\dot{M}_{\rm acc}$ and the millimeter luminosity $L_{\rm mm}$ could suggest different power laws, possibly associated with substructure.}
             }
    \label{fig:mdot_lmm}
\end{figure*}

\subsection{Evolutionary stage and substructure dependence in the $L_{\rm mm}$-$\dot{M}_{\rm acc}$ relation}\label{subsec:mdot_lmm}
The relationship between the mass accretion rate $\dot{M}_{\rm acc}$ and the millimeter luminosity $L_{\rm mm}$ provides an empirical link between stellar accretion and millimeter continuum emission from disks.
A positive $\dot{M}_{\rm acc}$-$M_{\rm dust}$ (corresponding $L_{\rm mm}$) correlation has been reported in several nearby star-forming regions, such as Lupus, Taurus-Auriga, and Corona Australis \citep{Manara_2016,Fiorellino_2022}.
We also detect a moderate positive correlation for all 54 disks in Ophiuchus (figure~\ref{fig:mdot_lmm}; $\rho=0.44$, the $p$-value=$2.7\times10^{-3}$).
Using \texttt{Linmix} \citep{Kelly_2007}, we perform Bayesian linear regression in logarithmic space and estimate the intercept $\gamma$ and the slope $\delta$ in the trend described as
\begin{align}
    \log L_{\rm mm} &= \gamma+\delta\log \dot{M}_{\rm acc}.
    \label{eq:mdot_lmm}
\end{align}
We obtain $\gamma=3.28\pm0.59$ and $\delta=0.22\pm0.07$, containing an intrinsic scatter of 0.55\,dex.
Given the large intrinsic scatter around the best-fit relation, we explore whether the $\dot{M}_{\rm acc}$-$L_{\rm mm}$ trend differs with evolutionary stage and disk substructures, as shown in figure~\ref{fig:mdot_lmm}.

Although the Class~I/FS subsample does not show a statistically significant correlation on its own (the $p$-value=0.08), both the Class~I/FS and Class~II subsamples exhibit moderate positive correlation coefficients. 
Moreover, the two populations largely overlap in the $L_{\rm mm}$-$\dot{M}_{\rm acc}$ plane (figure~\ref{fig:mdot_lmm}c). 
This contrasts with \citet{Fiorellino_2022}, who found that their Class~0/I sources are offset from the Class~II relation, with systematically higher mass accretion rates and disk masses. 
A plausible explanation is that our Class~I/FS sources are, on average, more evolved than the younger embedded Class~0/I objects studied by \citet{Fiorellino_2022}. 
This interpretation is supported by the relatively high bolometric temperatures of our sample ($T_{\rm bol}\sim$200-2000\,K) and by Paper~\citetalias{Shoshi_2025_oph}, which showed that the Ophiuchus sample is, on average, more evolved than the eDisk sample. 
If so, the evolutionary mismatch would naturally weaken the expected offset in the $L_{\rm mm}$-$\dot{M}_{\rm acc}$ relation, explaining the substantial overlap between our Class~I/FS and Class~II disks.

As shown in figure~\ref{fig:mdot_lmm}(a) and (b), disks with and without substructures occupy different regions in the $\dot{M}_{\rm acc}$-$L_{\rm mm}$ plane.
We apply Bayesian linear regressions with \texttt{Linmix} for disks without and with substructures to obtain $\gamma=3.25\pm0.99$ and $\delta=0.25\pm0.12$ with a standard deviation of 0.51 dex for the former and $\gamma=2.93\pm0.65$ and $\delta=0.13\pm0.08$ with a standard deviation of 0.48 dex for the latter.
The regression results indicate that though the inferred slopes are consistent within the uncertainties, the best-fit relations are slightly offset, such that some disks with substructures show higher $\dot{M}_{\rm acc}$ than smooth disks at a given $L_{\rm mm}$, which may indicate more dynamically active disks and/or more efficient inward mass transport.
Conversely, at fixed $\dot{M}_{\rm acc}$, substructured disks tend to have higher $L_{\rm mm}$ than smooth disks, suggesting that they retain a larger reservoir of dust.
This interpretation is qualitatively consistent with dust-trap scenarios, in which solids accumulate at pressure maxima \citep[][]{Youdin_2005,Johansen_2007,Johansen_2009}.
The distribution of substructured disks in the $\dot{M}_{\rm acc}$-$L_{\rm mm}$ plane is broadly consistent with the upper offset to smooth disks reported by \citet{Manara_2016} for transition disks in Lupus, which typically show single-ring morphologies.
Our results extend this result by showing that, in Ophiuchus, disks with substructures tend to be shifted toward higher $L_{\rm mm}$ at a given $\dot{M}_{\rm acc}$ relative to smooth disks, and that this tendency is not confined to transition disks but also appears in disks with a wider range of substructures (e.g., multiple rings and spirals).

At the same time, we note that $\dot{M}_{\rm acc}$ can vary by about an order of magnitude, likely reflecting the intrinsically time-variable nature of accretion from the inner disk onto the star \citep[e.g.,][]{Kenyon_1990,Safron_2015}.
In addition, our $\dot{M}_{\rm acc}$ values are compiled from multiple sources in the literature (table~\ref{table:properties}). 
The differences in the methods and assumptions to derive $\dot{M}_{\rm acc}$ may contribute to the observed scatter and introduce small systematic offsets.
We also tested whether replacing $L_{\rm mm}$ with an inclination-corrected quantity, $L_{\rm mm}/\cos i_{\rm disk}$, reduces the scatter in the $L_{\rm mm}$-$\dot{M}_{\rm acc}$ relation, as in the case of the $R_{95\%}$-$L_{\rm mm}$ relation.
However, the correlation does not become tighter.
A plausible explanation is that, unlike $R_{95\%}$, which is measured from deprojected radial profiles and is therefore already corrected for disk geometry to first order, the scatter in the $L_{\rm mm}$-$\dot{M}_{\rm acc}$ plane is likely dominated by intrinsic accretion variability and heterogeneous measurements rather than by inclination effects.
In that case, applying the inclination correction only to $L_{\rm mm}$ may introduce additional scatter rather than remove it.
Therefore, more homogeneous $\dot{M}_{\rm acc}$ measurements, ideally obtained over a more uniform range of observing epochs, would be valuable for rigorously assessing the robust relation of $L_{\rm mm}$-$\dot{M}_{\rm acc}$.

\begin{figure}[t]
    \begin{center}
        \includegraphics[width=\linewidth]{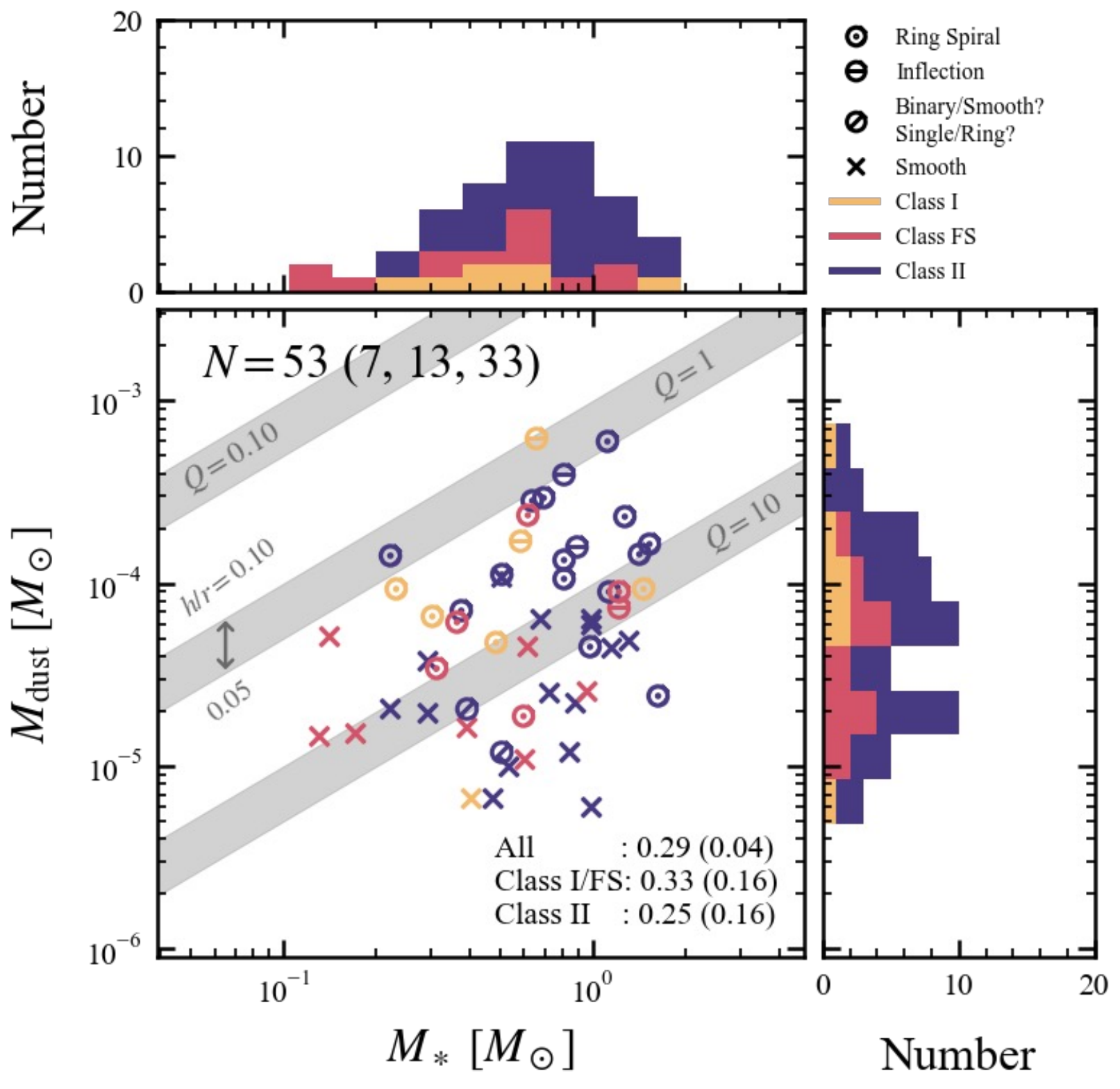}
    \end{center}
    \caption{
             Same as figure~\ref{fig:tbol_inc}, but for the relationship between stellar mass $M_\ast$ and dust disk mass $M_{\rm dust}$ for 54 Ophiuchus disks.
             The gray regions show the Toomre $Q$ parameter of 0.1 (upper), 1.0 (middle), and 10.0 (lower) derived by equation~\ref{eq:mdust_mstar}, where we assume the fixed values of a gas-to-dust mass ratio $\varepsilon$=100 and use the disk aspect ratio $h/r$ ranging from 0.05 to 0.10.
             {Alt text: The scatter diagram of stellar mass $M_\ast$ and dust disk mass $M_{\rm dust}$ suggests that most disks with substructures are distributed in the region with relatively large $M_{\rm dust}$.}
             }
    \label{fig:mstar_mdust}
\end{figure}

\begin{figure}[t]
    \begin{center}
        \includegraphics[width=\linewidth]{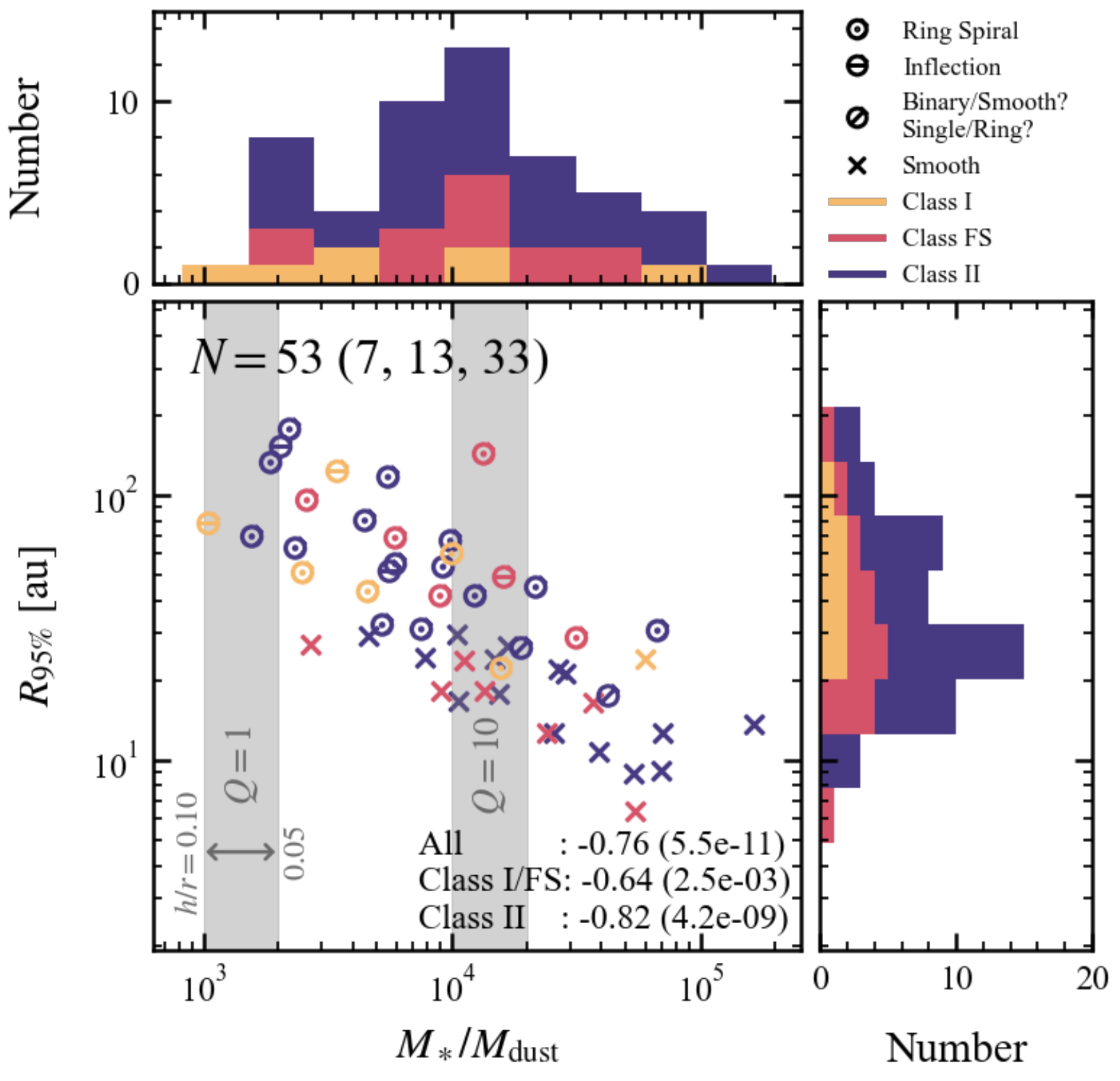}
    \end{center}
    \caption{
             Same as figure~\ref{fig:tbol_inc}, but for the relationship between the ratio of dust mass $M_{\rm dust}$ to stellar mass $M_\ast$ and dust radius $R_{95\%}$.
             The gray regions show the Toomre $Q$ parameter of 1.0 (left) and 10.0 (right) derived by equation~\ref{eq:mdust_mstar}, where we assume the fixed values of a gas-to-dust mass ratio $\varepsilon$=100 and use the disk aspect ratio $h/r$ ranging from 0.05 to 0.10.
             {Alt text: The scatter diagram between the ratio of dust mass $M_{\rm dust}$ to stellar mass $M_\ast$ and dust radius $R_{95\%}$ represents a clear anti-correlation, indicating that the scaling may be interpreted as $M_{\rm dust}$ increasing approximately with disk area ($M_{\rm dust}\propto R_{95\%}^2$).}
             }
    \label{fig:mstardust_radi}
\end{figure}   

\subsection{Possibility of gravitational instability}\label{subsec:lmm_mstar}
The size-luminosity relation $R_{95\%}$-$L_{\rm mm}$ in our sample suggests that the difference between substructured Class~II disks and the other populations, including Class~I/FS disks and smooth Class~II disks, could be explained by planet-induced disk evolution models such as those of \citet{Zormpas_2022}.
If the observed substructures are related to planet formation, it is important to assess whether the disk conditions are favorable for processes that promote planet formation.
One such process is classical gravitational instability (GI), which can operate in young, massive disks during the protostellar or accretion stage and may contribute to planet formation \citep[e.g.,][]{Durisen_2007,Kratter_2016,Tsukamoto_2023}.
Since disks with substructures in our sample tend to be bright and large (see section~\ref{subsec:corrsub}), and since the millimeter luminosity $L_{\rm mm}$ is used as a proxy for dust disk mass, we briefly investigate whether classical GI may play a role in individual disks in our sample.

The gravitational stability of a Keplerian disk can be assessed using the Toomre $Q$ parameter \citep{Toomre_1964},
\begin{align}
    Q &= \frac{c_s \Omega_{\rm K}}{\pi G \Sigma_{\rm gas}},\label{eq:tomreq}
\end{align}
where $c_s$ is the sound speed, $\Omega_{\rm K}$ is the Keplerian angular frequency, and $\Sigma_{\rm gas}$ is the gas surface density.
For a thin disk, the sound speed can be written as $c_s=h\Omega_{\rm K}$, where $h$ is the pressure scale height, so that the aspect ratio is $h/r$ at radius $r$.
Approximating the gas mass enclosed within radius $r$ as $M_{\rm gas} \simeq \pi r^2 \Sigma_{\rm gas}$ and adopting a constant gas-to-dust mass ratio $\varepsilon=100$ (i.e., $M_{\rm gas}=\varepsilon M_{\rm dust}$), equation~(\ref{eq:tomreq}) can be rewritten as
\begin{align}
    Q &= \frac{h}{r}\frac{M_\ast}{\varepsilon M_{\rm dust}},\label{eq:mdust_mstar}
\end{align}
which directly links the GI criterion to $M_\ast$ and $M_{\rm dust}$.

The dust disk masses $M_{\rm dust}$ of our sample are estimated from $L_{\rm mm}$ under the assumption of optically thin emission, with a constant dust temperature $T_{\rm dust}=20$\,K and a dust opacity coefficient $\kappa_\nu=\left(\nu/100\,{\rm GHz}\right)$\,cm$^2$g$^{-1}$ \citep[][]{Beckwith_1990}.
Following \citet{Williams_2019} and expressing the flux density $F_\nu$ and the distance $d$ in terms of the millimeter luminosity $L_{\rm mm}=F_\nu\times\left(d/140\right)^2$, the dust disk mass $M_{\rm dust}$ can be written as
\begin{align}
    M_{\rm dust}&=\frac{F_\nu d^2}{\kappa_\nu B_\nu\left(T_{\rm dust}\right)}=1.78\times10^{-6}\,M_\odot\left(\frac{L_{\rm mm}}{1\,{\rm mJy}}\right),\label{eq:mdust}
\end{align}
where $B_\nu$ is the Planck function.
Here, we note that the dust mass $M_{\rm dust}$ inferred by equation~(\ref{eq:mdust}) could be underestimated by factors of a few (and in some cases by orders of magnitude), primarily due to non-negligible optical depths and uncertainties in the effective dust opacity \citep[e.g.,][]{Liu_2022,Xin_2023}.

Figure~\ref{fig:mstar_mdust} shows the relationship between the stellar mass $M_\ast$ and the dust disk mass $M_{\rm dust}$ estimated from equation~(\ref{eq:mdust}).
For reference, we overplot the loci expected for $Q=0.1,\,1,\,10$ from equation~(\ref{eq:mdust_mstar}), assuming a gas-to-dust mass ratio of $\varepsilon=100$ and a representative range of disk aspect ratios $h/r=0.05$-0.10.
Although we cannot detect the flatter $M_{\rm dust}$-$M_\ast$ relation reported by \citet{Pinilla_2018} and \citet{Pinilla_2020}, disks with substructures occupy the relatively high-$M_{\rm dust}$ region of the diagram.
This tendency likely reflects the fact that disks with substructures tend to have higher $L_{\rm mm}$ (see section~\ref{subsec:corrsub}).
Furthermore, $M_{\rm dust}$ depends on the assumed dust opacity $\kappa_\nu$ and may be underestimated if the effective opacity is lower than adopted (and/or if the emission is partially optically thick).
As a result, the true locations of all the disks in the $M_\ast$-$M_{\rm dust}$ plane may shift toward higher $M_{\rm dust}$ and hence lower inferred $Q$.

Figure~\ref{fig:mstardust_radi} shows the relationship between the characteristic dust radius $R_{95\%}$ and the stellar-to-dust mass ratio $M_\ast/M_{\rm dust}$.
We use $M_\ast/M_{\rm dust}$ because this dimensionless quantity directly enters our order-of-magnitude estimate of the Toomre-$Q$ parameter (equation~\ref{eq:mdust_mstar}), allowing us to relate disk size to a simple measure of gravitational stability under the adopted assumptions.
The diagram shows a clear anti-correlation between $M_\ast/M_{\rm dust}$ and $R_{95\%}$: larger disks tend to have smaller $M_\ast/M_{\rm dust}$, i.e., larger inferred dust masses.
This behavior is consistent with the strong $R_{95\%}$-$L_{\rm mm}$ correlation (sections~\ref{subsec:corrsig} and \ref{subsec:lmm_r95}) and with our use of $L_{\rm mm}$ as a proxy for $M_{\rm dust}$.
Interpreted in terms of dust mass, the trend implies that $M_{\rm dust}$ increases with disk size and can be approximated as scaling with disk area, $M_{\rm dust}\propto R_{95\%}^2$.
Under our adopted $h/r$ and gas-to-dust ratio, several large, dust-rich disks can approach lower inferred $Q$ values, suggesting that classical GI may be relevant only to a limited subset of such disks.
However, the figure also includes many disks that remain in the gravitationally stable regime ($Q\gg1$), indicating that classical GI is unlikely to be relevant for most sources in our sample.

In summary, the comparison between $M_{\rm dust}$-$M_\ast$ and the Toomre $Q$ parameter suggests that most disks in our sample remain in the gravitationally stable regime, and that only a limited subset of large, dust-rich disks may approach conditions under which classical GI becomes relevant.
This interpretation should be treated with caution because the inferred $Q$ values depend on the assumed gas-to-dust mass ratio, disk aspect ratio, dust opacity, and optical depth.
To place firmer constraints on whether classical GI plays any role in individual disks, future work will require independent estimates of the gas surface density from molecular-line observations, together with multi-wavelength continuum data and radiative transfer modeling to account for dust optical depth and opacity uncertainties \citep[][]{Tazzari_2021,Miotello_2018}.
It will also be important to examine whether other processes, such as secular GI \citep[e.g.,][]{Takahashi_2016,Tominaga_2020,Tominaga_2023}, can promote dust concentration and planet formation in disks where classical GI is unlikely to operate.

\begin{figure*}[ht]
    \begin{center}
        \includegraphics[width=\linewidth]{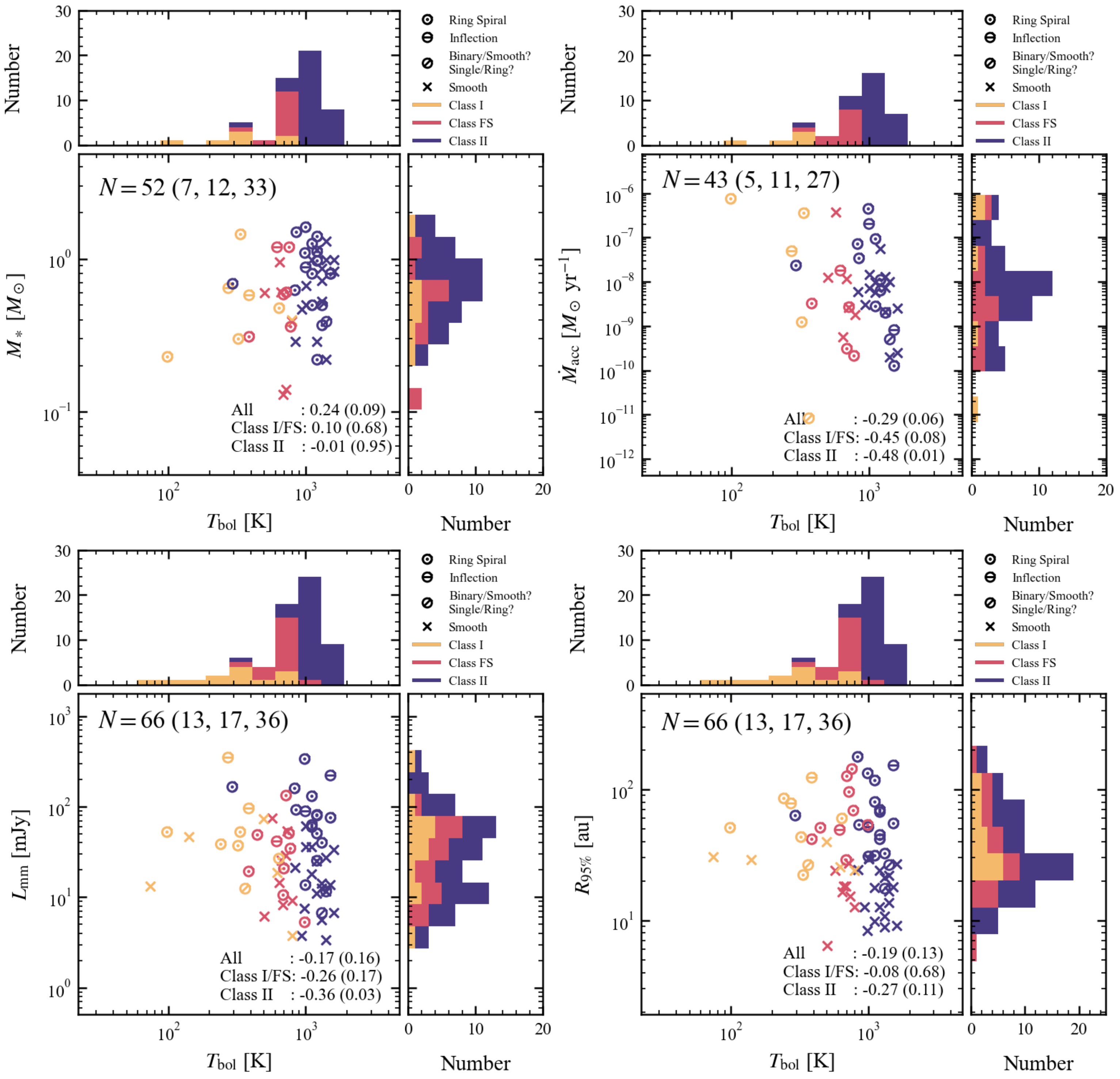}
    \end{center}
    \caption{
             Relationships of bolometric temperature $T_{\rm bol}$ with other stellar and disk properties $M_\ast$ (upper left), $\dot{M}_{\rm acc}$ (upper right), $L_{\rm mm}$ (lower left), and $R_{95\%}$ (lower right).
             The colors of the symbols indicate the evolutionary stage classified by the spectral slope at 2-22\,$\mu$m, where yellow, red, and violet correspond to Class~I, FS, and Class~II stages.
             The marker shapes denote the disk categorizations; $\odot$ as ``Ring'' or ``Spiral'', $\ominus$ as ``Inflection'', $\oslash$ as candidates for nearly edge-on disks with ``Ring'' features or circumstellar disks in binary systems, and $\times$  as ``Smooth'' brightness distributions.
             The values at the lower right of the scatter diagram show the Pearson correlation coefficients $\rho$, with the corresponding $p$-values in brackets, for all disks, Class~I and FS disks, and Class~II disks.
             The top panel shows the histogram of $T_{\rm bol}$, and the right panel shows that of each property, indicating the sample size for each quantity.
             {Alt text: The scatter diagrams show the relations for bolometric temperature $T_{\rm bol}$ with other stellar and disk properties $M_\ast$, $\dot{M}_{\rm acc}$, $L_{\rm mm}$, and $R_{95\%}$.}
             }
    \label{fig:tbol_scatters}
\end{figure*}

\begin{figure*}[ht]
    \begin{center}
        \includegraphics[width=\linewidth]{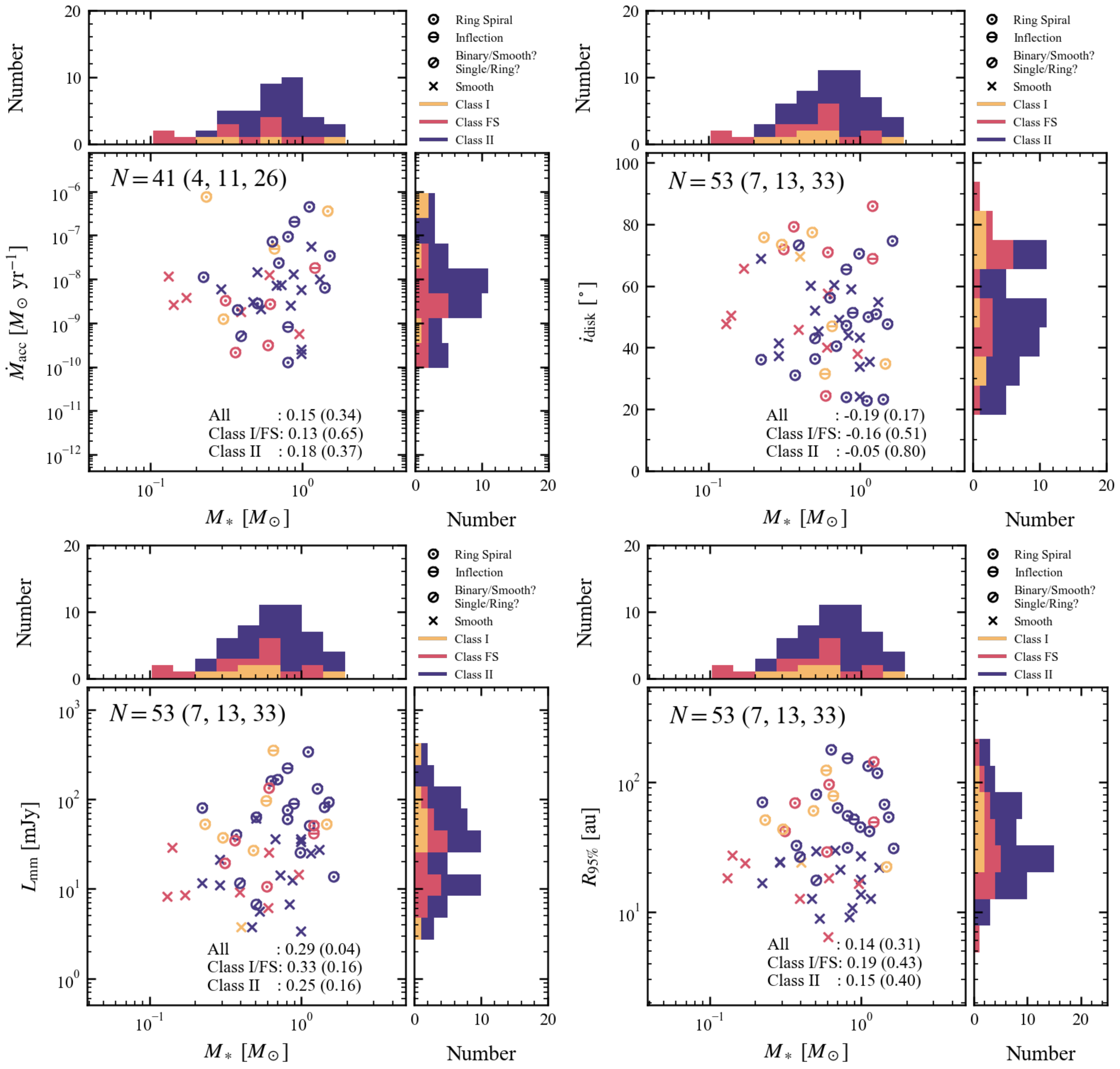}
    \end{center}
    \caption{
             Same as Figures~\ref{fig:tbol_scatters} but for the relationships of stellar mass $M_\ast$ with other stellar and disk properties $\dot{M}_{\rm acc}$ (upper left), $i_{\rm disk}$ (upper right), $L_{\rm mm}$ (lower left), and $R_{95\%}$ (lower right).
             {Alt text: The scatter diagrams show the relations for stellar mass $M_\ast$ with other stellar and disk properties $\dot{M}_{\rm acc}$, $i_{\rm disk}$, $L_{\rm mm}$, and $R_{95\%}$.}
             }
    \label{fig:mstar_scatters}
\end{figure*}

\begin{figure*}[ht]
    \begin{center}
        \includegraphics[width=\linewidth]{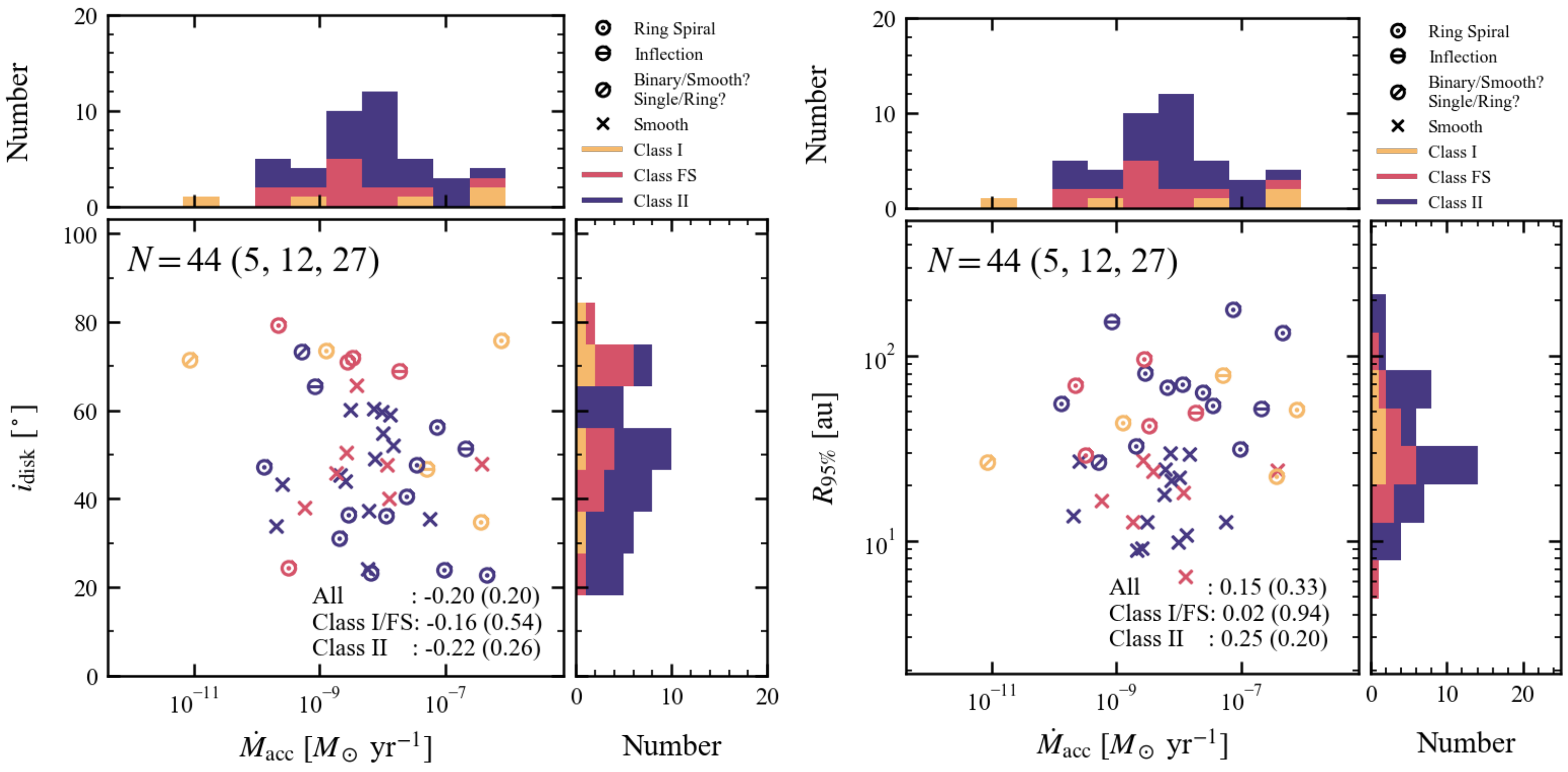}
    \end{center}
    \caption{
             Same as Figures~\ref{fig:tbol_scatters} but for the relationships of mass accretion rate $\dot{M}_{\rm acc}$ with other disk properties $i_{\rm disk}$ (left) and $R_{95\%}$ (right).
             {Alt text: The scatter diagrams shows the relations for mass accretion rate $\dot{M}_{\rm acc}$ with other disk properties $i_{\rm disk}$ and $R_{95\%}$.}
             }
    \label{fig:mdot_scatters}
\end{figure*}

\begin{figure*}[ht]
    \begin{center}
        \includegraphics[width=\linewidth]{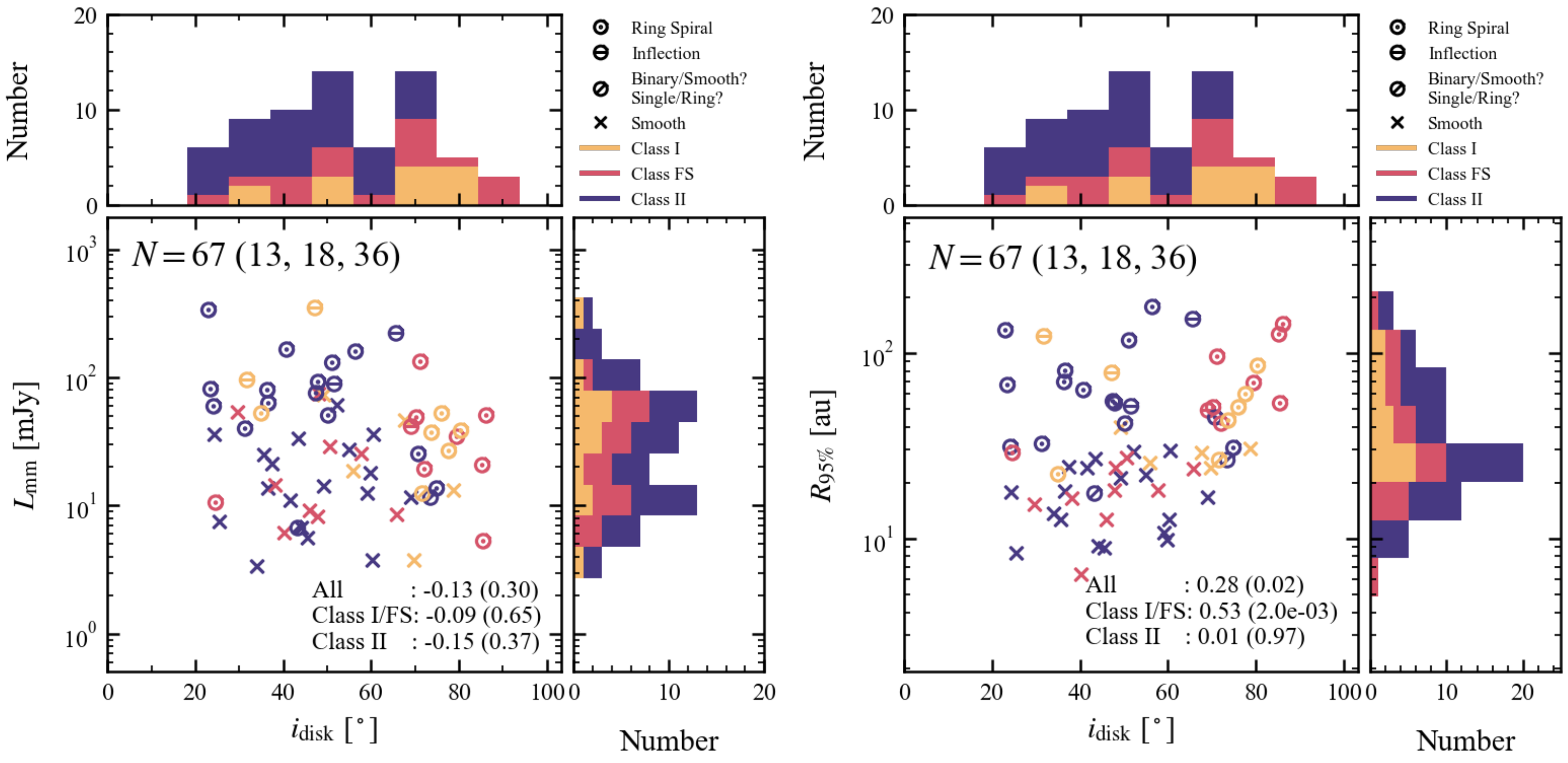}
    \end{center}
    \caption{
             Same as Figures~\ref{fig:tbol_scatters} but for the relationships of inclination angle $i_{\rm disk}$ with other disk properties $L_{\rm mm}$ (left) and $R_{95\%}$ (right).
             {Alt text: The scatter diagrams show the relations for inclination angle $i_{\rm disk}$ with other disk properties $L_{\rm mm}$ and $R_{95\%}$.}
             }
    \label{fig:inc_scatters}
\end{figure*}

%%% Summary %%%
\section{Summary}\label{sec:summary}
We present a statistical analysis of stellar and spatially resolved dust disk properties for young stellar objects in the Ophiuchus star-forming region.
Based on the 78 disks analyzed with PRIISM two-dimensional super-resolution imaging in Paper~\citetalias{Shoshi_2025_oph}, we define a conservative sample of 67 systems for which dust radii and related disk properties can be robustly measured.
By combining stellar parameters from the literature with disk properties derived from PRIISM images, we investigate how correlations among these quantities vary with evolutionary stage and the presence of disk substructures.

Our main results are summarized as follows:
\begin{enumerate}
    \item
    Only a limited number of parameter pairs show robust or tentative correlations in the Ophiuchus sample.
    The most robust trend is the $R_{95\%}$--$L_{\rm mm}$ relation, while the $L_{\rm mm}$--$\dot{M}_{\rm acc}$, $L_{\rm mm}$--$M_\ast$, and $T_{\rm bol}$--$i_{\rm disk}$ correlations remain tentative.
    Most other parameter pairs are consistent with scatter.
    
    \item
    Detectable substructures are more strongly associated with disk properties than with stellar properties.
    Substructured disks are systematically brighter and larger in $L_{\rm mm}$ and $R_{95\%}$ than smooth disks, whereas no significant differences are found in $T_{\rm bol}$, $M_\ast$, or $\dot{M}_{\rm acc}$ within the current sample.
    The tentative difference in $i_{\rm disk}$ is not physically meaningful because $i_{\rm disk}$ is a viewing-angle parameter, and substructure detectability depends on the viewing geometry.
    
    \item
    Although the full sample shows a weak anti-correlation between $T_{\rm bol}$ and $i_{\rm disk}$, this trend is not reproduced within the evolutionary subsamples.
    We therefore interpret it primarily as an effect of sample incompleteness rather than a strong physical inclination bias.
    Overall, $T_{\rm bol}$ remains a relatively robust statistical tracer of evolutionary stage in this sample.
    
    \item
    A key result of this work is the tight $R_{95\%}$-$L_{\rm mm}$ relation, enabled by the homogeneous disk radii measured from PRIISM super-resolution images.
    Only Class~II disks with substructures show a distinctly steeper relation ($\beta\sim0.8$), whereas the other subsamples remain broadly consistent with $\beta\sim0.4$-0.5.
    This behavior is consistent with the evolutionary models of \citet{Zormpas_2022}, in which disks with planet-induced pressure bumps evolve toward a steeper size--luminosity relation than smooth disks.
    
    \item
    We find a moderate positive $L_{\rm mm}$-$\dot{M}_{\rm acc}$ correlation in the full sample.
    Smooth and substructured disks have similar slopes, but substructured disks tend to be brighter at a given $\dot{M}_{\rm acc}$, suggesting more efficient dust retention.
    This interpretation is plausible but remains uncertain due to the heterogeneity in measurements of $\dot{M}_{\rm acc}$ in the literature.
    
    \item
    The comparison between $M_{\rm dust}$-$M_\ast$ and the Toomre $Q$ parameter suggests that most disks in our sample are gravitationally stable, and that classical GI is likely relevant only for a limited subset of large, dust-rich disks.
    Future studies should therefore investigate whether other processes can promote dust concentration and planet formation in these systems.
    However, this result remains tentative due to uncertainties in dust mass estimates, disk aspect ratios, and turbulence properties.
\end{enumerate}

We note that our resolution and sensitivity cuts may under-sample compact and faint disks, while observational biases in the detectability of substructures may also influence the inferred trends.
In addition, uncertainties in optical depth and dust opacity, as well as heterogeneous assumptions in the compiled literature values for stellar parameters, may contribute to the observed scatter.
Multi-wavelength continuum and spatially resolved molecular-line observations will be essential to constrain disk masses and gas surface densities better, thereby clarifying the physical origin of the reported trends.

%%% Acknowledgement %%%
\begin{ack}
The authors thank the anonymous referee for all of the comments and advice that helped improve the manuscript and the contents of this study.
This work is part of the ASIAA Summer Student Program 2023, and I appreciate the support of the Academia Sinica Institute of Astronomy and Astrophysics.
The authors thank Dr. T. Nakazato and Dr. S. Ikeda for their technical support.
This work was supported by a NAOJ ALMA Scientific Research grant (No. 2022-22B; MNM) and by JSPS KAKENHI JP25KJ1947 (AS), JP26K17220 (MY), JP26K00741 (MY), JP23K20872 (TT), JP24K07097 (TT), JP23K03463 (TM), and JP25K07369 (MNM). 
This study uses the following ALMA data: ADS/JAO.ALMA \#2016.1.00545.S. 
ALMA is a partnership of ESO (representing its member states), NSF (USA) and NINS (Japan), together with NRC (Canada), MOST and ASIAA (Taiwan), and KASI (Republic of Korea), in cooperation with the Republic of Chile. 
The Joint ALMA Observatory is operated by ESO, AUI/NRAO and NAOJ.
This work presents results from the European Space Agency (ESA) space mission Gaia. Gaia data are being processed by the Gaia Data Processing and Analysis Consortium (DPAC). Funding for the DPAC is provided by national institutions, in particular, the institutions participating in the Gaia MultiLateral Agreement (MLA). The Gaia mission website is $\langle$\url{https://www.cosmos.esa.int/gaia}$\rangle$. The Gaia archive website is $\langle$\url{https://archives.esac.esa.int/gaia}$\rangle$.
This research has made use of the VizieR catalogue access tool, CDS, Strasbourg, France (DOI : 10.26093/cds/vizier).
The original description of the VizieR service was published in 2000, A\&AS 143, 23.
Data analysis was carried out on the Multi-wavelength Data Analysis System operated by the Astronomy Data Center (ADC), National Astronomical Observatory of Japan.
We acknowledge the use of OpenAI's ChatGPT only as a grammar-checking and editing tool to improve the clarity and readability of the manuscript.
This paper made use of the following software: AnalysisUtilities $\langle$\url{https://casaguides.nrao.edu/index.php?title=Analysis_Utilities}$\rangle$, Astropy \citep{astropy_2022}, CASA \citep{CASA_2022}, Linmix \citep{Kelly_2007}, matplotlib \citep{Hunter_2007}, PRIISM \citep{Nakazato_2020,Nakazato_2020b}, SciPy \citep{Virtanen_2020}, and NumPy \citep{Harris_2020}.
\end{ack}

%%% Appendix %%%
\appendix
\section{All scatter diagrams}\label{sec:other_scatter}
Figure~\ref {fig:matrix} represents the correlations of 15 possible pairs among the six kinds of stellar and dust disk properties, bolometric temperature $T_{\rm bol}$, stellar mass $M_\ast$, mass accretion rate $\dot{M}_{\rm acc}$, inclination angle $i_{\rm disk}$, millimeter luminosity $L_{\rm mm}$, and dust disk radius $R_{95\%}$.
In section~\ref{sec:discussion}, we discuss $i_{\rm disk}$-$T_{\rm bol}$ on the misclassification for evolutionary stages, $M_{\rm dust}$-$M_\ast$ on the possibility of GI, and the two combinations,  $L_{\rm mm}$-$\dot{M}_{\rm acc}$ and $R_{95\%}$-$L_{\rm mm}$, on the dependence of disk substructures.
Therefore, the other pairs, including $L_{\rm mm}$-$M_\ast$ which is the original of $M_{\rm dust}$-$M_\ast$, are shown in figures~\ref{fig:tbol_scatters}-\ref{fig:inc_scatters}.
The colors and shapes of markers in figures~\ref{fig:tbol_scatters}-\ref{fig:inc_scatters} denote the evolutionary stages of the systems and the categorizations of the disks.

\begin{figure}[t]
    \begin{center}
        \includegraphics[width=\linewidth]{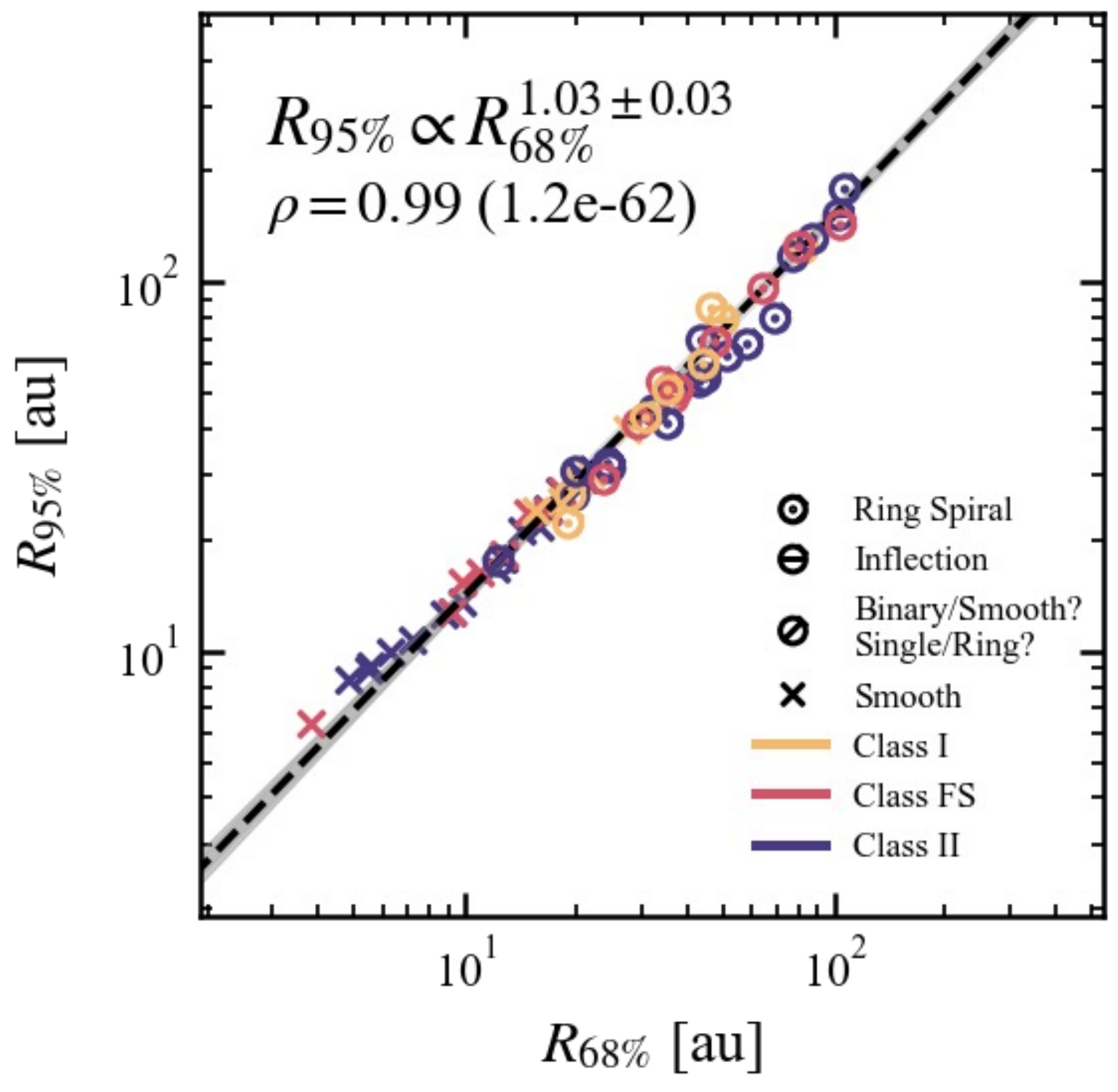}
    \end{center}
    \caption{
             Comparison of dust disk radii enclosing 68\% and 95\% of the total flux density $R_{68\%}$ and $R_{\rm 95\%}$.
             The colors of the symbols indicate the evolutionary stage classified by the spectral slope at 2-22\,$\mu$m, where yellow, red, and violet correspond to Class~I, FS, and Class~II stages.
             The marker shapes denote the disk categorizations; $\odot$ as ``Ring'' or ``Spiral'', $\ominus$ as ``Inflection'', $\oslash$ as candidates for nearly edge-on disks with ``Ring'' features or circumstellar disks in binary systems, and $\times$  as ``Smooth'' brightness distributions.
             The black dashed line indicates the median scaling relation obtained from Bayesian linear regression, and the dark gray shaded region represents the 68\% confidence interval around it. 
             The best-fitting linear regression parameters, the Pearson correlation coefficient $\rho$, and the associated $p$-value are listed at the upper right corner.
             {Alt text:The scatter diagram shows the linear relationship between dust radii $R_{68\%}$ and $R_{95\%}$.}
             }
    \label{fig:r68r95}
\end{figure}

\section{Dust disk radii defined by enclosed flux}\label{sec:r68r95}
In the literature, the dust disk radius has been defined using different enclosed fractions of the total continuum flux. 
\citet{Tripathi_2017} and \citet{Andrews_2018_sma}, based on low- to intermediate angular resolution SMA and ALMA observations, adopted the radius enclosing 68\% of the total flux, $R_{68\%}$, as a representative disk size. 
This definition has also been widely used in numerical studies \citep[e.g.,][]{Rosotti_2019,Zormpas_2022}. 
However, high-angular-resolution ALMA observations have demonstrated that $R_{68\%}$ does not always capture the full radial extent of disks and can miss some substructures, such as rings and gaps \citep[e.g.,][]{Long_2019,Huang_2018}. 
Consequently, more recent studies have increasingly used dust radii defined at higher enclosed-flux fractions, including $R_{90\%}$ and $R_{95\%}$.

To facilitate a consistent comparison with previous studies, we investigate the relationship between $R_{68\%}$ and $R_{95\%}$ measured in Paper~\citetalias{Shoshi_2025_oph}.
Figure~\ref{fig:r68r95} shows the relation between $R_{68\%}$ and $R_{95\%}$ for 67 disks in Ophiuchus. 
The two radii are strongly correlated, with a Pearson correlation coefficient $\rho$ close to unity and a very small $p$-value, indicating a statistically significant correlation. 
We further apply Bayesian linear regression and obtain
\begin{align}
    \log\,\left(\frac{R_{95\%}}{1\,{\rm au}}\right)=\left(0.12^{+0.05}_{-0.05}\right)+\left(1.03^{+0.03}_{-0.03}\right)\log\,\left(\frac{R_{68\%}}{1\,{\rm au}}\right)
\end{align}
with an intrinsic scatter of 0.03\,dex. 
This tight relation implies that $R_{95\%}$ can be reliably inferred from $R_{68\%}$, enabling disk sizes reported under different enclosed-flux definitions to be placed on a common scale.

%%% Reference %%%
\makeatletter
\renewcommand{\bibsection}{\par\vspace{12pt}\noindent{\bibsectionfont\refname}\par\vspace{-3pt}}
\makeatother
\bibliographystyle{forbibtex}
\bibliography{refpaper}
\end{document}